\documentclass[graybox]{svmult}

\usepackage{mathptmx}       % selects Times Roman as basic font
\usepackage{helvet}         % selects Helvetica as sans-serif font
\usepackage{courier}        % selects Courier as typewriter font
\usepackage{type1cm}        % activate if the above 3 fonts are
\usepackage{makeidx}         % allows index generation
\usepackage{graphicx}        % standard LaTeX graphics tool
\usepackage{multicol}        % used for the two-column index
\usepackage{url}             % allows inclusion of URLs
\usepackage[bottom]{footmisc}% places footnotes at page bottom
\usepackage[authoryear,comma]{natbib}
\usepackage{float, amsmath, amssymb, pdfsync}
\usepackage{amsbsy,bm}
\makeindex             % used for the subject index
\begin{document}
\title*{Pulsar Wind Nebulae as Cosmic Pevatrons: A Current Sheet's Tale$^*$}

\author{
Jonathan Arons
}

\institute{
Department of Astronomy, Department of Physics, Space Sciences Laboratory and Theoretical Astrophysics Center \\ University of California, Berkeley\\ 
\email{arons@berkeley.edu} \\
\newline
\hspace*{-0.03in} *With apologies to Geoffrey Chaucer and Margaret Atwood 
}

\maketitle

%\abstract{
%Each chapter should be preceded by an abstract (10--15 lines long) that summarizes the content. The abstract will appear \textit{online} at \url{www.SpringerLink.com} and be available with unrestricted access. This allows unregistered users to read the abstract as a teaser for the complete chapter. As a general rule the abstracts will not appear in the printed version of your book unless it is the style of your particular book or that of the series to which your book belongs.
%Please use the 'starred' version of the new Springer \texttt{abstract} command for typesetting the text of the online abstracts (cf. source file of this chapter template \texttt{abstract}) and include them with the source files of your manuscript. Use the plain \texttt{abstract} command if the abstract is also to appear in the printed version of the book.
%}

\abstract{I outline, from a theoretical and somewhat personal perspective, significant features of Pulsar Wind Nebulae as Cosmic Accelerators.  I discuss recent studies of Pulsar Wind Nebulae (PWNe). I pay special attention to the recently discovered gamma ray ``flares'' in the Crab Nebula's emission, focusing on the possibility, raised by the observations, that the accelerating electric field exceeds the magnetic field, suggesting that reconnection in the persistent current layer (a ``current sheet'') plays a significant role in the behavior of this well studied Pevatron.  I address the present status of the termination shock model for the particle accelerator that converts the wind flow energy to the observed non thermal particle spectra, concluding that it has a number of major difficulties related to the transverse magnetic geometry of the shock wave. I discuss recent work on the inferred pair outflow rates, which are in excess of those predicted by existing theories of pair creation, and use those results to point out that the consequent mass loading of the wind reduces the wind's bulk flow 4-velocity to the point that dissipation of the magnetic field in a pulsar's wind upstream of the termination shock is restored to life as a viable model for the solution of the ``$\sigma$'' problem.  I discuss some suggestions that current starvation in the current flow supporting the structured (``striped'') upstream magnetic field, perhaps inducing a transition to superluminal wave propagation. I show that current starvation probably does not occur, because those currents are carried in the current sheet separating there stripes rather than in the stripes themselves\renewcommand{\thefootnote}{\fnsymbol{footnote}}
\footnote[2]{Collaborators, none of whom should be held responsible for the content of this paper: D. Alsop, E. Amato, D. Backer, P. Chang, N. Bucciantini, B. Gaensler, Y. Gallant, V. Kaspi, A.B. Langdon, C. Max, E. Quataert, A. Spitkovsky, M. Tavani,  A. Timokhin}}. 
\renewcommand{\thefootnote}{\arabic{footnote}}
%\vspace*{1cm}

\section{Gamma Ray Flares from the Crab Nebula: Observations \label{sec:flare_obs}}

\begin{figure}[H]
\begin{center}
%\unitlength = 0.0011\textwidth
\hspace{10\unitlength}
\begin{picture}(300,100)(0,15)
\put(0,-75){\makebox(300,200)[tl]{\includegraphics[width=3.5in]{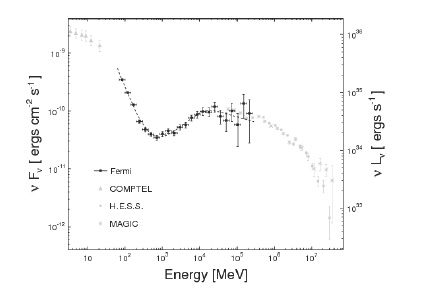}}}
%\put(200,-100){\makebox(300,200)[tl]{\includegraphics[width=4in]{./flareSpectra2009-10.pdf}}}
\end{picture}
\end{center}
\vspace{1.5cm}
\caption{Gamma ray ($10 \; {\rm TeV} > \varepsilon > 1$ MeV) average spectrum of the Crab Nebula in 2008-11, from \citet{buehler11}.} 
\label{fig:average}
\end{figure}

Rotation Powered Pulsars (RPPs) and their Pulsar Wind Nebulae (PWNe) are cosmic accelerators containing the highest energy particles in identified sources.  Modeling photon emission, which is synchrotron radiation from radio ($\mu$eV) up to GeV gamma rays, provides evidence for electrons and (probably) positrons with energies up to a few PeV, as has been known for many years. In the last two years, the new generation of satellite borne gamma ray telescopes, AGILE and FERMI, have discovered strong variability in the $\varepsilon > 100$ MeV gamma ray flux from the Crab Nebula that may revise, or at least extend, ideas about the acceleration of such very high energy particles in well-identified sources - the sources of the ultra-high energy cosmic rays, whose observed flux contains particles with energy up to $\sim 0.1 $ ZeV, have yet to be identified \citep{kotera10}. Figure \ref{fig:average} shows the average spectrum of the Crab seen by the Large Area Telescope (LAT) in the first 33 months  after the launch of the Fermi satellite.

%\vspace*{0.25in}

The standard model for the high energy emission employs some form of acceleration of non-thermal distributions of electrons (and positrons) at the termination ``shock'' of the relativistic, magnetized radiatively inefficient (``cold'') wind that carries the central pulsar's rotational energy loss from the magnetosphere (dimension $\sim R_L =c/\Omega_* = 1576 (P/33 \; {\rm msec})$ km (the light cylinder distance) to the location of the ``termination working surface\footnote{I prefer this somewhat old fashioned term for the surface or region where outflow energy converts to non-thermal heat, since as will be seen, the interpretation of the surface as a traditional MHD shock wave does not work, requiring one to endow the word "shock" with unusual properties, if it is to be used in this context.}'' (the TWS) at $R_{TWS} = (\dot{E}_R/4\pi c P_{nebula})^{1/2} = 3.6 \times 10^{12}$ km =0.4 Light years = 0.12 pc $= 2.3 \times 10^9 R_L$.  For the Crab Pulsar, timing observations of the pulsar'' period and period derivative yield $\dot{E}_R = -I \Omega_* \dot{\Omega}_* = 5 \times 10^{38}$ ergs/s, with a factor $\sim 2$ uncertainty due to uncertainty in the magnitude of the neutron star's moment of inertia $I$, itself due to imperfect knowledge of the equation of state of the nuclear matter in the star's interior. 

\begin{figure}[H]
\begin{center}
%\unitlength = 0.0011\textwidth
\hspace{10\unitlength}
\begin{picture}(300,100)(0,15)
\put(0,-270){\makebox(300,300)[tl]{\includegraphics[width=4in,height=4.5in]{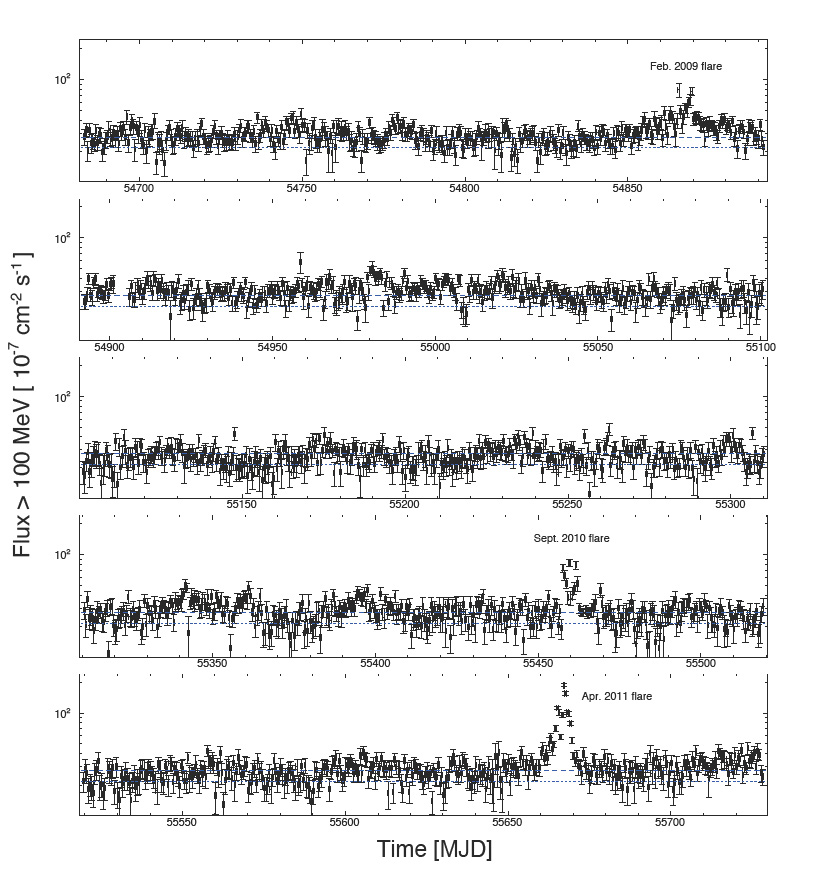}}}
\end{picture}
\end{center}
\hspace*{0.1cm}
\vspace*{10cm}
\caption{Gamma ray ($\varepsilon > 100$ MeV)  light curve of the Crab Nebula as detected by the Fermi LAT, for the thirty-three months ending in April 2011, binned in 12  hour intervals \citep{buehler11}. The flare of September-October 2007, detected by AGILE before Fermi's launch \citep{tavani11}, is not shown here.}  
\label{fig:thirty-threeMonths}
\end{figure}

\begin{figure}[H]
\begin{center}
%\unitlength = 0.0011\textwidth
\hspace{10\unitlength}
\begin{picture}(300,100)(0,15)
\put(0,-105){\makebox(300,200)[tl]{\includegraphics[width=4in]{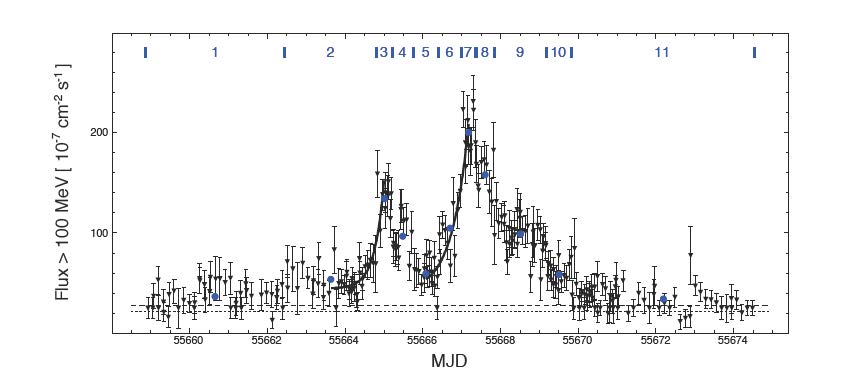}}}
\end{picture}
\end{center}
\vspace{1.5cm}
\caption{Gamma ray ($\varepsilon > 100$ MeV) April 2011 flare light curve, from \citet{buehler11}, showing the fast (few hours) variations .} 
\label{fig:LightCurves2011}
\end{figure}

\begin{figure}[H]
\begin{center}
%\unitlength = 0.0011\textwidth
\hspace{10\unitlength}
\begin{picture}(300,100)(0,15)
%\put(0,-55){\makebox(300,200)[tl]{\includegraphics[width=4in]{./April2011LightCurve.pdf}}}
\put(0,-15){\makebox(300,200)[tl]{\includegraphics[width=4in]{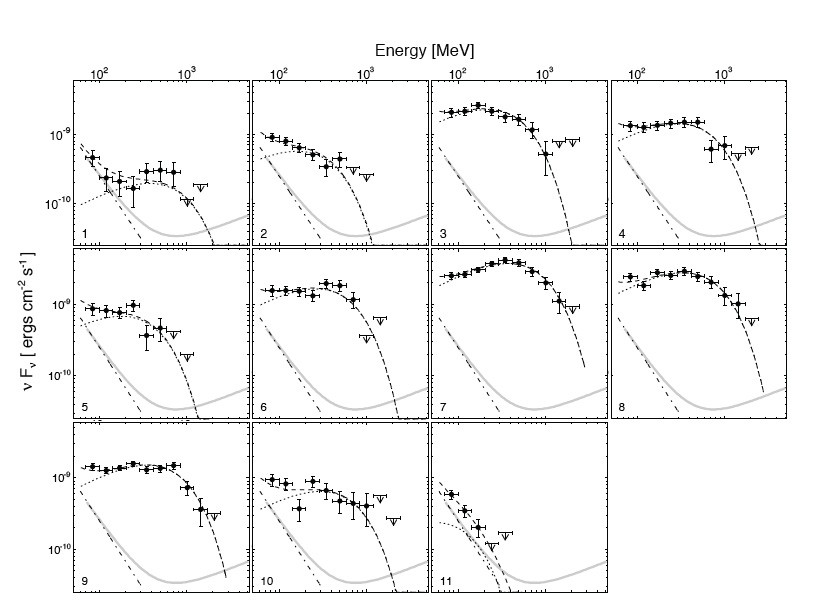}}}
\end{picture}
\end{center}
\vspace{1.5cm}
\caption{Gamma ray ($\varepsilon > 100$ MeV) April 2011 flare spectra at 11 times during the flare, from \citet{buehler11}. The time windows are
indicated by the numbers in the bottom left corner of each panel and correspond to the intervals shown in
Figure \ref{fig:LightCurves2011}. The
dotted line shows the SED of the flaring component (logarithmic flux $\varepsilon F_\varepsilon$ in ergs cm$^{-2}$ sec$^{-1}$as a function of photon energy $\varepsilon$), the dot-dashed line the constant background from the
synchrotron nebula, and the dashed line is the sum of both components. The average Crab nebular
spectrum in the first 33 months of Fermi observations is also shown in gray for comparison. Superficially, these spectra suggest the acceleration of an almost monoenergetic component of the radiating particles (a ``beam''?), with this unusual accelerator persisting for a few days, or, if lasting longer, beamed toward us for only a few days.} 
\label{fig:spectra2011flare}
\end{figure}

Figure \ref{fig:thirty-threeMonths}  shows the Nebula's light curve above 100 MeV using 12 hour averages during the same period.  Apparent are the restless small amplitude variability and the three flares (February 2009, September 2010 and April 2011) seen by AGILE and FERMI.  Figure \ref{fig:LightCurves2011} shows the same kind of data for the most spectacular event observed in April 2011. During that event, the peak power reached 1\% of the pulsar's spin down power (the ultimate source of the energy), thirty times the average gamma ray brightness at the same energy, with factors of a few flux variation on time scales of hours \citep{balbo11}.  The 2011 flare spectrum drastically hardened compared to the normal, average nebular gamma ray spectrum, as seen in Figure \ref{fig:spectra2011flare}.  Despite the restless behavior of the $\gamma$-ray emission at all times, the events identified as flares do seem to be special events - unlike the largest earthquakes\footnote{see \cite{langer94} for a review of earthquake dynamics.}, they are not the tail of a (non-Gaussian) distribution of amplitudes (J. Scargle, 2011, personal communication). Throughout all these events, the pulsed emission coming directly from the pulsar did not change - no variations in the pulsar spin aside from the regular spin down have been detected, and no changes in the spectra or light curves of the pulsed emission appeared (\citealt{abdo11}, \citealt{buehler11}). Thus, the blame for these phenomena has to be assigned to events in the nebula, the calorimeter which makes luminous the otherwise dark outflow from the central compact object.

The wind model for the excitation of the Crab Nebula, introduced in a form mixed with the vacuum wave model for pulsar spin down by \citet{rees74} and re-introduced in its fully MHD form by \citet{kennel84}, is designed to quantity the steady emission from the Crab Nebula and other PWNe\footnote{The relativistic MHD wind model for pulsar spin down was introduced by \citet{michel69}}.  It has been shown to incorporate unsteady, quasi-periodic variability with time scale $\sim$ a year \citep{spit04, camus09} $\sim$ flow transit time from pulsar to the TWS and in the post TWS flow, with simulated behavior rather akin to the variable ``wisp'' features long known from optical \citep{lampland21, scargle69}, X-ray \citep{weisskopf00} and radio \citep{biet01, biet04} studies. The wisp variability time scale's similarity to the time separating the well demarcated flares perhaps suggests the origin of the gamma ray flares lies in the physics of the TWS.  As it stands, however, that model has no ready explanation for the short variability time during a flare - the hierarchy of time scales (many months between flares, few days flare duration, and few hours intra-flare variability) has not been part of the modeling to date. In addition,  the spectral hardening, looking as if a new, transient component of the PeV particle distributions appeared, then vanished, poses a serious challenge to the standard model. I discuss the significance of these results in \S \ref{sec:flares-analysis}
\newpage
\vspace*{-1.75in}
\vspace*{-0.6in}
%\vspace*{1in}
\newpage
\vspace*{1cm}
%\vspace*{-0.3in}

\section{Pulsar Wind Nebulae: Basic Concepts and the Standard Model}
  
\begin{figure}[H]
\begin{center}
%\unitlength = 0.0011\textwidth
\hspace{10\unitlength}
\begin{picture}(300,100)(0,15)
\put(0,-100){\makebox(300,200)[tl]{\includegraphics[width=4in]{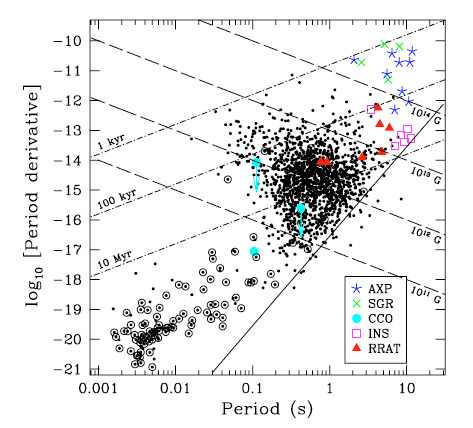}}}
\end{picture}
\end{center}
\vspace{7.5cm}
\caption{Observed RPP periods and period derivatives, adapted from \cite{kaspi10}.   The line bounding the pulsar population corresponds to the magnetospheric voltage $\Phi_{\rm mag} = 10^{12}$ V. It clearly marks a boundary beyond which pulsar emission is unlikely.  Simple estimates of pair creation  within the magnetosphere, which assume the existence of a parallel voltage drop in the strong field region $r < R_L$ with magnitude a large fraction (generally greater than a few \%) of $\Phi_{\rm mag}$, suggests this source of plasma should occur only for $\Phi_{\rm mag} > 10^{12}$ V. The numerical value is determined by the pair conversion opacity of photons propagating in the super strong $B$ field and the magnetic geometry; see \citet{hibsch01} and \citet{medin10} for modern evaluations. This result underpins the idea that pair creation is essential for radio emission \citep{sturrock71}. The change in slope of this ``death line'' at short period, small $\dot{P}$ indicates more sophistication in the pair creation physics, and/or in the association of pairs with radio emission, than is incorporated in the simplest models, a conclusion also apparent from the variety of quantitative problems with this widely accepted hypothesis \citep{hibsch01, medin10}.}
 \label{fig:PPdot}
\end{figure}

PWNe and their generative pulsars provide the first known and most definitive example of compact astrophysical systems which draw the power for their observed emissions from the electromagnetic extraction of rotational energy from  gravitationally bound objects.  They have motivated models for similar energy extraction from disks around other gravitating bodies, such as black holes (e.g. \citealt{rees84, begelman84}). As objects of study, the RPPs have a virtue lacking in other systems studied in high energy astrophysics: timing of the precisely measured pulse periods, uniquely interpretable as the rotation periods of the underlying neutron stars, provide measurements of the total energy budget free of all astrophysical ambiguities, other than the factor of $\sim 2$ uncertainty in neutron stars' moments of inertia, arising from the lack of precise knowledge in the equation of state of dense matter. The measured rate of rotational energy loss, 
\begin{equation}
\dot{E}_R = -I \Omega_* \dot{\Omega}_* = 4\pi I \frac{\dot{P}}{P^3},
\label{eq:spinloss}
\end{equation}
tells us the total energy budget for these systems, without our having to understand anything about the photon emissions. The observed distributions of $P, \dot{P}$ appea in Figure \ref{fig:PPdot}. This is both a blessing and a curse - a blessing, because in contrast to other relativistic astrophysical systems, we know the total luminosity in all forms, seen and unseen, without having to unravel the partition between flow kinetic energy, large scale Poynting flux, thermal energy and radiative losses - a curse, because the energy loss is radiatively silent, thus supplying little information as to the details of the energy outflow, leaving the mechanics of the machine mysterious.  

\subsection{Magnetospheres}

Nevertheless, progress has been made on interpreting the physics of the observed radiation. The most significant advances have come in interpreting the Pulsar Wind Nebulae (PWNe), since these act as catch basins for the rotational energy lost.  Observations of these systems %\citet{gaensler06}, using radio (including millimeter), X-ray, gamma ray and occasionally infrared telescopes\footnote{Most PWNe lie in the galactic plane, therefore are relatively inaccessible to optical techniques, and are even less accessible to UV telescopes.  Near and far infrared observations are extremely useful in unraveling the physics of the relativistic outflows \citep{bucc10}, but have been much less in evidence than the high energy studies.}, 
have made clear that RPPs deliver their energy to the outside world in the form of highly relativistic, magnetized outflows - stellar winds that are exaggerated versions of the solar wind - which must be electromagnetically driven by the magnetic pressure of the wound up magnetic field. The strength of that field is estimated by using the theory of magnetic braking of the neutron stars' spin, which suggests 
\begin{equation}
\dot{\Omega}  =  -K \Omega^n, \; \Omega = 2\pi /P, 
\label{eq:spindown}
\end{equation}
applied to the observed rotation periods and spindown rates $\dot{P} = - 2\pi \dot{\Omega}/\Omega^2$.

The earliest model applied vacuum electrodynamics to a rotating sphere endowed with a magnetic dipole moment $\bm{\mu}$ centered at the stars' centers and tipped with respect to the rotation axis by an angle $i$.  That theory yields the spindown luminosity $\dot{E}_R = K \Omega^4, \; K = (2/3)\mu^2 \sin^2 i /c^3$  \citep{pacini67, ostriker69}. Thus $n = 3$ in this model, if $\mu$ and $i$ are constant. Vacuum theory was motivated by the large gravitational forces at the surfaces, suggesting no plasma more than a meter or so above the star \citep{hoyle64}.  But immediately the much earlier observation \citep{deutsch55}, made in the context of magnetic A stars, that in vacuum large electric fields parallel to $B$ would overwhelm gravity and pull charged particles out from the star until the vacuum electric field would be altered, reducing $\bm{E} \cdot \bm{B}$ down to zero, was recovered \citep{gold69} and extended with the suggestion that the charged particles would feed a curious charge separated wind, with a total electric current $I = c \Phi_{\rm mag}, \; \Phi_{\rm mag} = $ total magnetospheric potential $= \sqrt{\dot{E}_R /c} = 4 \times 10^{16} (\dot{P}/3 \times 10{-13})^{1/2} (33 \; {\rm msec} /P^3)^{3/2}$ Volts - the numerical result is normalized to the Crab pulsar.  That wind could carry away the rotational energy of even the aligned rotator, in a Poynting flux dominated flow - the electromagnetic energy density would vastly exceed the kinetic energy density (and pressure) of the outflow, in the initially conceived model - the particle flux in that scheme is only $c\Phi/e = 2.3 \times 10^{30} (I_{45} \dot{P}_{15}/P^3)^{1/2} $ elementary charges/s, $I_{45} = I/10^{45} \; {\rm cgs}, \dot{P}_{15} =  \dot{P}/10^{-15}$, the ``Goldreich-Julian'' current. 

The charge separated model has several really serious theoretical difficulties, but perhaps of greater importance is that the observations of {\it young} PWNe ($\Phi_{\rm mag} > 10^{14.5}$ Volts) have made  clear that the particle outflows are many orders of magnitude larger than the Goldreich-Julian current, thus motivating MHD models, where ${\boldsymbol E} \cdot {\boldsymbol B} = 0$ is assumed from the start. Pair creation occurring somewhere within the magnetospheres (first suggested by \citealt{sturrock71}) is thought to be the origin of the required dense plasma, although a quantitative model that yields the high mass loss rates observed is still lacking \citep[and references therein.]{bucc11}  In recent years, solutions for the MHD structure of the magnetosphere in the appropriate force-free limit have been obtained (numerically - attempts of analytically minded theorists to guess the detailed answer uniformly failed over 30 years of trying), first for the aligned rotator \citep{contop99, gruzinov05, komiss06, timokhin06}, then for the full 3D oblique rotator \citep{spit06, kala09}.

\begin{figure}[H]
\begin{center}
%\unitlength = 0.0011\textwidth
\hspace{10\unitlength}
\begin{picture}(300,100)(0,15)
\put(0,-100){\makebox(300,200)[tl]{\includegraphics[width=4in]{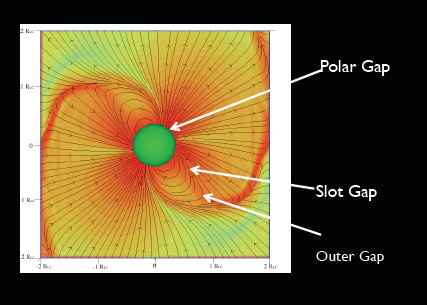}}}
\end{picture}
\end{center}
\vspace{5.8cm}
\caption{Field lines and current density of the oblique force-free rotator, $i = 60^\circ$, from Spitkovsky (2006), seen in a cut in the ${\boldsymbol{\Omega}, \boldsymbol{\mu}}$ plane. The last closed field lines end at a Y-point (a Y-line, in 3D) at distance $R_Y$ from the neutron star -- technically, the closed zone ends at cusp, not a Y-line, since the separatrix between the closed and open zones interior to the line of closure carries the return current \citealt{uzdensky97}, but this distinction is of no importance to the present discussion. The current sheet encloses and separates the closed from the open field regions of the magnetosphere, and the separate branches merge in the wind zone, where the folded sheet continues to separate the oppositely directed fields of the striped wind. The arrows and ``Gap'' labels locate sites where vacuum gaps have been postulated, in test particle  models of accelerators that lead to gamma ray emission and pair creation. None of these gaps appear in a current sheet accelerator based on the force-free magnetosphere model.}
 \label{fig:currentsheet}
\end{figure}

As far as the $P, \; \dot{P}$ diagram goes, the main inference from modeling the force-free magnetosphere is the innocuous looking result for the scale factor $K$ in expression (\ref{eq:spindown}): 
\begin{equation}
K = k(1 + \sin^2 i) \frac{\Omega^3 \mu^2}{c^3}, \;  k = 1 \pm 0.1.
\label{eq:spindownnorm}
\end{equation}
Physically, the most important result has been the identification of the current sheets separating the closed from the open regions of the magnetosphere, extending into the wind beyond the magnetosphere, whose last closed flux surface ends just touching the light cylinder whose cylindrical radius is $\varpi = c/\Omega $.  That such a current sheet should be present has been suspected from the early days of RPP research ({\it e.g.,} \citealt{michel75}). The error in $k$ reflects the uncertainties in the numerical treatment of the problem, many of which are associated with how the current sheet is represented in the numerical schemes. Figure \ref{fig:currentsheet} shows a slice through the  3d force-free magnetosphere of the $60^\circ$ rotator, from Spitkovsky's (2006) results. 
%\vspace*{-1cm}
\noindent

\begin{figure}[H]
\begin{center}
%\unitlength = 0.0011\textwidth
\hspace{10\unitlength}
\begin{picture}(300,110)(0,15)
\put(0,-150.00){\makebox(300,200)[tl]{\includegraphics[width=4in,height=3in]{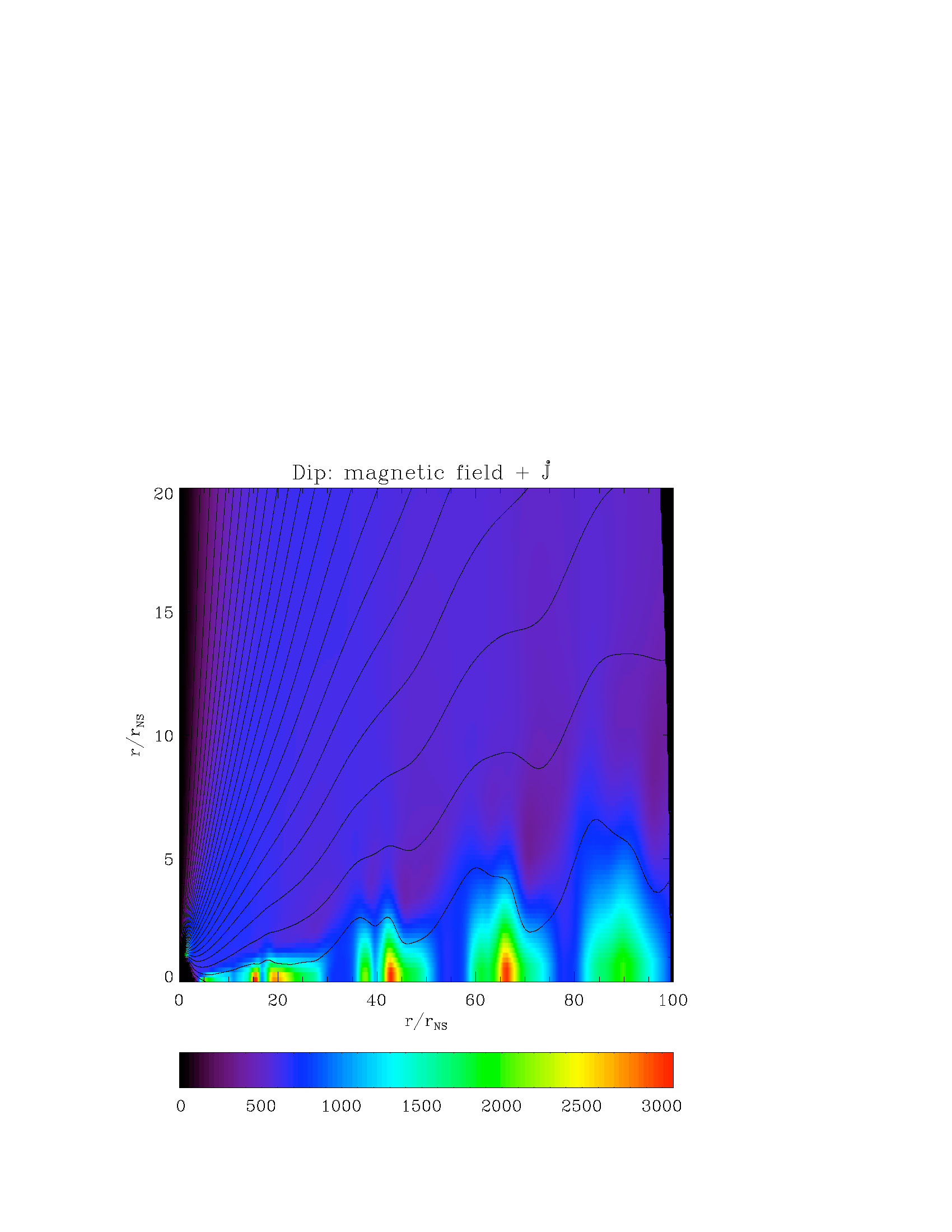}}}
\end{picture}
\end{center}
\vspace{5.8cm}
\caption{Plasmoids formed at the base of the current sheet of the relativistic aligned rotator, where the sheet crosses the light cylinder, from a relativistic MHD simulation of a newly born neutron star's magnetosphere \citep{bucc06}.  The dissipation that allows reconnection to occur is numerical. They move out radially at the local Alfven speed $v_A \approx c$, and recur on the magnetospheric Alfven transit time $\sim P /\pi$}
 \label{fig:plasmoids}
\end{figure}

As far as the $P, \; \dot{P}$ diagram goes, the main inference from modeling the force-free magnetosphere is the innocuous looking result for the scale factor $K$ in expression (\ref{eq:spindown}): 
\begin{equation}
K = k(1 + \sin^2 i) \frac{\Omega^3 \mu^2}{c^3}, \;  k = 1 \pm 0.1.
\label{eq:spindownnorm}
\end{equation}
Physically, the most important result has been the identification of the current sheets separating the closed from the open regions of the magnetosphere, extending into the wind beyond the magnetosphere, whose last closed flux surface ends just touching the light cylinder whose cylindrical radius is $\varpi = c/\Omega $.  That such a current sheet should be present has been suspected from the early days of RPP research ({\it e.g.,} \citealt{michel75}). The error in $k$ reflects the uncertainties in the numerical treatment of the problem, many of which are associated with how the current sheet is represented in the numerical schemes. Figure \ref{fig:currentsheet} shows a slice through the  3d force-free magnetosphere of the $60^\circ$ rotator, from Spitkovsky's (2006) results.  
Observationally, expression (\ref{eq:spindownnorm}) shows that the vacuum rotator's ``braking index'' $n = 3$ is preserved in full force-free MHD, which contradicts the observed values in the small number of stars where $n$ has been determined \citep{livingstone07}, a contradiction which has led to a variety of suggestions ranging from evolution of the magnetic moment $\mu$ or the obliquity $i$ \citep{blandford88} to effects of reconnection on the rate of conversion of open magnetic flux to closed \citep{contop06}, as the star spins down and the closed zone expands at the expense of the amount of open magnetic flux. That reconnection might affect the braking index is readily derived from the fact that the torque really depends on the magnitude of the open magnetic flux.The amount of open flux depends on the size of the closed zone, which ends at $R_Y$.  If $R_Y/R_L < 1$, the torque increases because of more open field lines and larger Poynting  flux than is the case for a magnetosphere closing at $r = R_L$. \cite{bucc06} show that the braking index is 
\begin{equation}
n \equiv \frac{\Omega \ddot{\Omega}}{\dot{\Omega}^2} = 
    3 + 2 \frac{\partial \ln \left(1 + \frac{R_L}{R_Y} \right ) } {\partial \ln \Omega};
\label{eq:braking_recc}
\end{equation}
thus, If $R_Y/R_L$ decreases with decreasing $\Omega$, then $n<3$.  Reconnection usually is unsteady - Figure \ref{fig:plasmoids} shows the blobs (``plasmoids'') ejected from the Y-line at the  %\vspace*{-1cm}
base of the current sheet in the relativistic wind (beyond the light cylinder) of the aligned rotator. 
\vspace*{-2.75cm}
The speculation is that spindown slightly biases these reconnection events so that on the much longer spindown timescale, the net open flux slowly converts to closed. These current fluctuations might be associated with the timing ``noise'' \citep{arons81a, cheng87} identified long ago with torque  fluctuations ({\it e.g.}, \citealt{helfand80, scott03}), although recent analysis of longer data sets \citep{lyne10} in long period pulsars has called the pure noise interpretation into question.

In itself, the force-free model does not provide mechanisms for photon emission.  But it has a variety of implications, which are slowly being addressed.

\begin{itemize}
\item The model specifies the polar flux tube size and shape - for $i \neq 0$, it is noncircular with a polar cap center displaced from the magnetic axis, even when the magnetic field is the simplest, that of a star centered, point dipole \citep{bai10}. This has consequences for radio polarization structure, and for polar cap areas and dipole offsets inferred from soft X-ray emission from polar caps \citep{bogdanov07}, thought to be heated by magnetospheric particle bombardment \citep{arons81b, zavlin98, harding02}. These theoretical improvements of the polar cap model have yet to be noticed and incorporated in phenomenological models of the observations used by data analysts; such incorporation might yield interesting tests of the force-free model. \citet{chung11} show how ``Stokes Tomography'' applied to radio polarization might be usefully employed in this task.

\item The force free model quantitatively specifies the return currents required to prevent the star from charging up, as the polar flow extracts charge from the star. The results for the oblique rotator are partly in accord with long held expectations, that in part return current exists in a thin (``auroral'') sheet bounding the polar flux tube ({\it e.g.} \citealt{gold69, michel75}), consistent with the open circuited model \citep{gold69} - current closure occurs far away, in the nebula/interstellar medium beyond the wind termination shock or perhaps in the outer wind - plus qualitatively new features: a) part of the return current surrounding the polar flux tube is spatially distributed, even in the aligned rotator; b) in the oblique rotator, part of this current system is not a return current at all, but couples the two polar regions together \citep{bai10} - in the orthogonal rotator ($i = 90^\circ$), the auroral component of the current is entirely in the polar coupling flow, with the volume current out of each half of the polar cap having equal amount and opposite sign, also consistent with early expectations ({\it e.g.} \citealt{saf78}) - the orthogonal rotator automatically balances its charge loss. A preliminary account of some aspects of ``auroral'' particle acceleration based on the force-free model are outlined in \citet{arons11}. The radiative consequences of these features are as yet unexplored - for example, the spatially distributed part of the return current might be a good candidate for the site of ``conal'' component of pulsar radio emission, an idea which requires non-force-free modeling of the current flow and identification of a workable emission process within that current flow model\footnote{Some earlier ideas on this subject relating to field aligned acceleration and gamma ray emission can be found in \citet{arons83b, gruzinov08}, for example.} before one can relate the theoretical force free current distributions to the observations in a testable manner (although easier kinematic comparisons are certainly possible).
\item The location of the return current layer having been determined, the hypothesis that the return current layer is the site of the beamed particle accelerator that gives rise to the pulsed gamma rays observed by the FERMI and earlier orbiting gamma ray telescopes (see \citealt{ray11}) can now be investigated in the context of a self consistent magnetospheric structure that allows a quantitative evaluation of the beaming characteristics implied by the radiating current sheet concept - see \citet{bai10} for a kinematical study of the radiating current sheet idea.

\end{itemize} 

\subsection{Current Sheet and Pulsed $\gamma$-ray emission in the Lighthouse Model: Relativistic Aurorae}

\begin{figure}[H]
\begin{center}
%\unitlength = 0.0011\textwidth
\hspace{10\unitlength}
\begin{picture}(300,110)(0,15)
\put(0,-75){\makebox(300,200)[tl]{\includegraphics[width=4in]{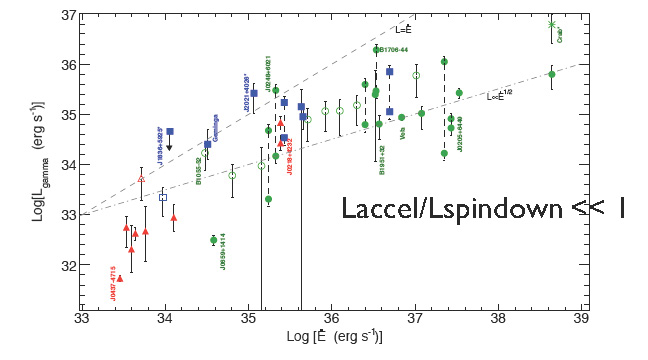}}}
\end{picture}
\end{center}
\vspace{1cm}
\caption{Ratio of observed gamma ray luminosity $L_{\gamma}$ to observed spindown power $\dot{E}_R = L_{\rm spindown}$ for LAT pulsars \citep{abdo10a}.  If the acceleration of the gamma ray emitting particles is radiation reaction limited (generally true in models that assume curvature radiation as the emission mechanism, as is the case in most of the ``gap'' models), $L_\gamma $ is a good proxy for the acceleration power.  The trend at higher luminosities roughly follows the efficiency $L_{\gamma}/L_{\rm spindown} \propto \dot{E}_R^{-1/2}$ expected if some mechanism limits the accelerating voltage $\Delta \Phi$ to a fixed fraction of the magnetospheric voltage $\Phi_{\rm mag} = \sqrt{\dot{E_R}/c}$, ({\it e.g.}, \citealt{arons96, harding02}), a limitation usually (and plausibly) attributed to pair creation.  The large dispersion in Figure \ref{fig:GammaEffic} comes primarily from the uncertain distances, and also from the uncertain beaming correction required to go from the fraction of the sky illuminated by the beam to the total radiative -- the LAT team assumed uniform phase-averaged  beaming across the sky (1 sterradian).}
 \label{fig:GammaEffic}
\end{figure}

Making progress on a physical model for radiation from the current layers can most expeditiously take advantage of the facts that a) pulsed gamma ray emission, when observed, is the largest photon output from rotation powered pulsars, but b)  generally has less luminosity than the spin-down luminosity of these stars. Figure \ref{fig:GammaEffic} illustrates this fact, which summarizes the results from the LAT instrument as of Spring 2010. Thus the energy invested in particle acceleration is a small fraction of the energy stored in Poynting fluxes, for radiation reaction limited acceleration, so that the force free model can be considered as a good zeroth order magnetospheric description.

The voltage limitation prediction $L_{\rm gamma} \propto \Phi_{\rm mag} \propto \sqrt{\dot{E}}$ is the same as $L_{\gamma} \propto$ ``particle current'' = particle flux = Goldreich-Julian flux in unidirectional beam models of the polar electric current flow \citep{harding81} only if the  accelerator carries a fixed fraction of the total electric current that enters into the spin-down torque, independent of $P, \dot{P}$.  In the traditional slot or outer gap models, pair creation establishes the limitation of the accelerator to a thin sheet either in the outer magnetosphere \citep{ruderman86} or back in the polar cap \citep{arons83a}. In such models, the gap width $w_*$ of a model that successfully reproduces the sharply peaked light curves, projected onto the neutron star following the field lines, is necessarily small compared to the polar cap size, \emph{and} varies with the pulsars' magnetic moment and spin parameters. Thus $L_\gamma \propto \Phi_{\rm mag}$ is not the same as $L_{\gamma} \propto$ electric current $I = c\Phi_{\rm mag}$.  This is not an issue for a model based on the currents flowing in the return current layer, which necessarily carries the whole magnetospheric return current, for most obliquities of the magnetic moment with respect to the rotation axis.

\begin{figure}[H]
\begin{center}
%\unitlength = 0.0011\textwidth
\hspace{10\unitlength}
\begin{picture}(300,110)(0,15)
\put(50,-150){\makebox(300,200)[tl]{\includegraphics[width=3in]{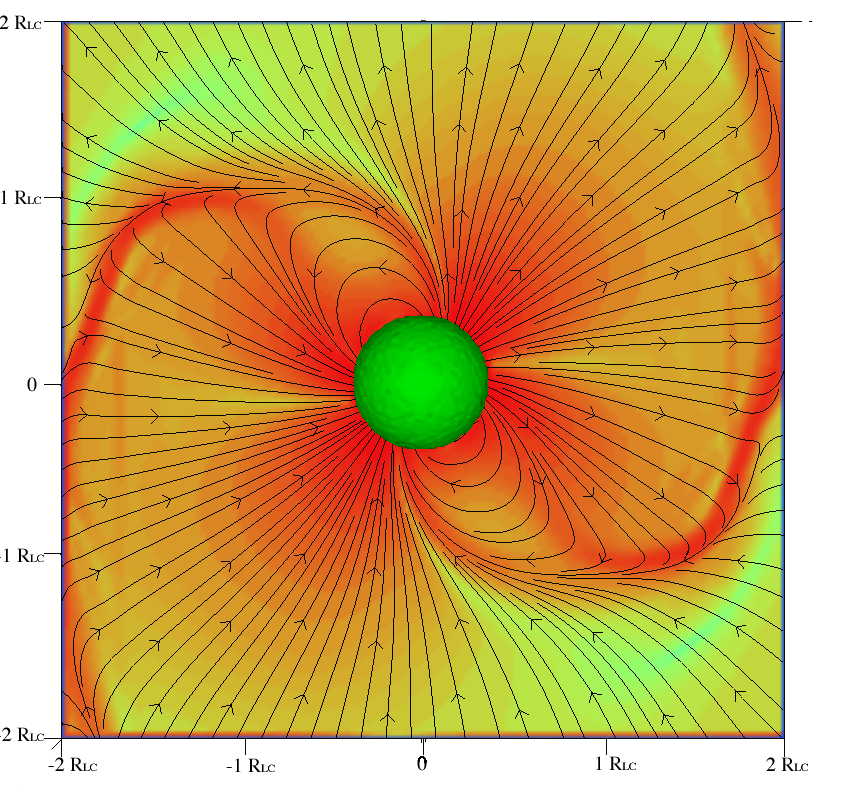}}}
\end{picture}
\begin{picture}(300,-110)(0,15)
\put(-20,-350){\makebox(300,200)[tl]{\includegraphics[width=4.65in]{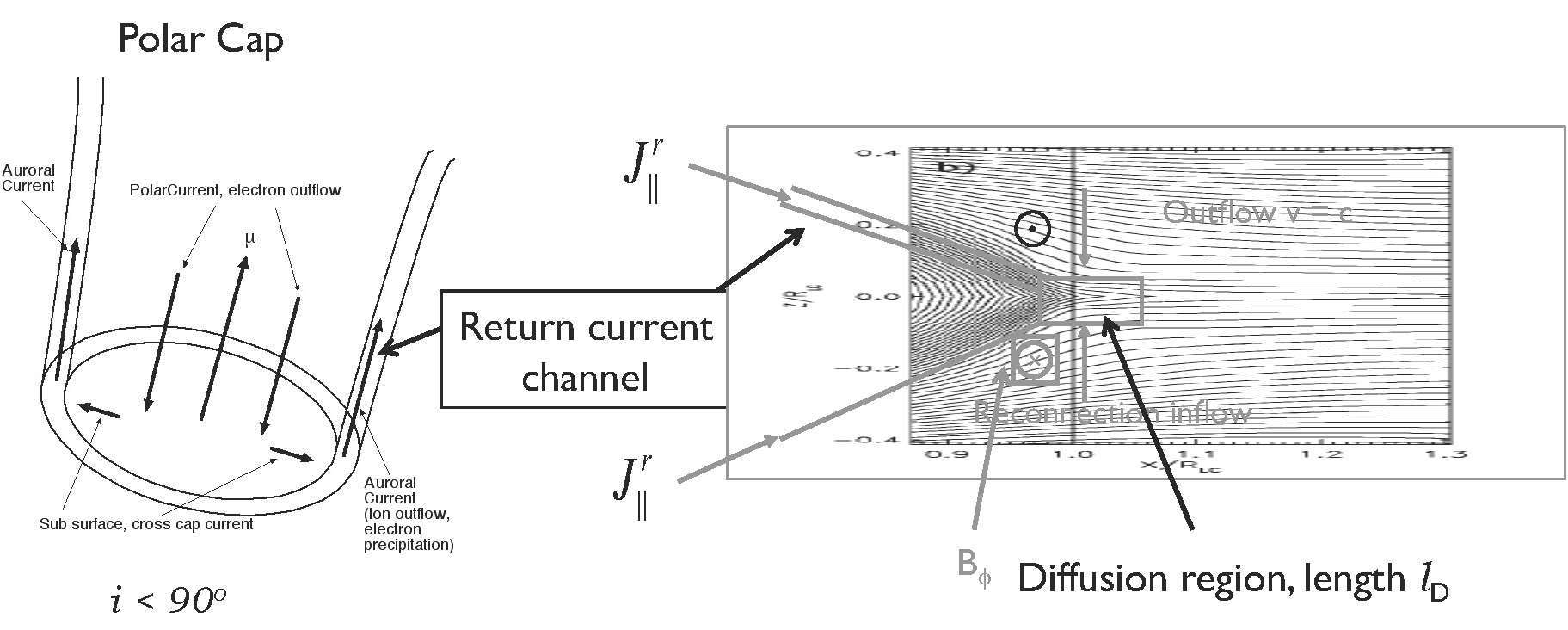}}}
\end{picture}
\end{center}
\vspace{10.5cm}
\caption{Upper Panel:  Electric current structure of the oblique force-free magnetosphere, inclination = $60^\circ$, from \cite{spit06}. The current sheet, indicated by the darker color, bounds the closed zone. The closed zone ends at a singular Y-line (in the ideal force-free electrodynamics approximation) at the light cylinder distance from the neutron star. Lower Panel, left: Current structure at the polar cap, illustrated for acute angle between the magnetic moment and the angular velocity (electron polar current, ${\boldsymbol \Omega} \cdot {\boldsymbol \mu} > 0$).  The return current in the current sheet consists of a precipitating electron beam, launched from the diffusion region around the Y-line and possibly augmented by high altitude pair creation within the current sheet, with the charges in the precipitating beam extracted by reconnection flow from the pair plasma flowing from the polar cap into the wind, plus a counterstreaming ion beam extracted electrostatically from the stellar atmosphere. In obtuse geometry (ion polar current, ${\boldsymbol \Omega} \cdot {\boldsymbol \mu} < 0$), positrons precipitate from the Y-line and counterstreaming electrons are extracted from the atmosphere. For clarity, the part of the return current not contained in the current sheet is omitted, even though this part of the current system is of increasing significance as $i \rightarrow 90^\circ$. Lower Panel, right: Possible structure of the Y-line region, with the termination of the closed zone to the left and the merger of the winds from the two polar caps to the right.  The ``guide field'' $B_\phi$ also reverses across the mid-plane of the flow, along with the poloidal open field.  As reconnection occurs, some of the ouflowing plasma (speed $c\beta, \; \beta \approx 1$) deflects toward the singular, unmagnetized ``diffusion'' region around the Y-line with speed $v_{rec} \sim 0.1 v_{\rm Alfven} = 0.1 c$. The figure represents a steady (in the co-rotating frame) flow model - in reality, the reconnection is likely to be bursty, as in Figure \ref{fig:plasmoids}, with formation of sporadic X-lines.}
 \label{fig:current_system}
\end{figure}

That magnetospheric current sheets, with particle densities within the sheets high compared to the Goldreich-Julian value, can sustain large parallel electric fields is well known in planetary magnetospheres - such sheets are the accelerator sites of the particle beams that stimulate the aurora observed in the upper atmospheres of the Earth, Jupiter and Saturn, for example\footnote{For a review of such phenomena, see \cite{pasch02}. Of particular significance to pulsars is the fact that the field aligned currents that power narrow auroral arcs consist of precipitating electron beams launched from the reconnection region in the distant magnetotail and counterstreaming ions launched from the planetary atmosphere.}  An elementary illustration of this possibility comes from considering the inertial term in the generalized Ohm's for the electric field parallel to $\boldsymbol{B}$, which can be written as, in the relativistic case, as
\begin{equation}
E_\parallel = \frac{4\pi}{\omega_p^2}
     \left\{ 
      \frac{\partial J_\parallel}{\partial t} +
     \frac{{\boldsymbol {\vec B}}}{B} \cdot {\boldsymbol {\vec \nabla}} \cdot 
        \left[ \gamma 
        \left( {\boldsymbol {\vec J} {\vec v}} + {\boldsymbol {\vec v}{\vec J}} \right )
         \right] 
         \right\}
         \propto
         \frac{m\gamma}{n} \frac{I}{\Delta_{\rm current} \rho_B},
\label{eq:genOhm}
\end{equation}
with $m \gamma$ the particles' relativistic mass, $n$ their density, $I$ the total current set by the force free magnetosphere, $\Delta_{\rm current}$ is the thickness of the current carrying channel, and $\rho_B$ is the radius of curvature of the magnetic field. 
%\vspace*{-1.5cm}
\noindent Thus inertia of the current carriers can act as a effective resistance in these high inductance systems, which establish the currents electromagnetically. The presence of this ``resistor'' in the circuit forces a parallel electric field to appear.  The same is true of any parallel (to $B$) load - pressure and radiation reaction are other effects  which can serve as loads, with pressure especially important in the diffusion region around the singular line shown in figure \ref{fig:current_system}. Equation (\ref{eq:genOhm}) is most useful in the corotating frame, and when the current is due to relative motion between the species (electrons and positrons, and in some circumstances heavy ions) that is slow compared to the bulk fluid velocity {\bf v}.  In the current circumstance, it turns out that the current is better described as counter-streaming beams - in that case, describing the beams as separate fluids is more appropriate, and they can form the total plasma density in the current flow channel, rather than being a low density component in a much denser plasma.

Expression (\ref{eq:genOhm}) does make clear that acceleration is prone to maximize when the  the relativistic mass is high and the current density is high (large $I/\Delta_{\rm current})$. $\Delta_{\rm current}$ is generally microscopic, expected to be on the order of $c/\omega_p$, and is established by the dynamics of the the singular region where the closed zone ends, illustrated (in the cartoon approximation) in Figure \ref{fig:current_system}. The capture rate of pair plasma into the diffusion region is
\begin{equation}
\dot{N}^{in}_{{\rm diffusion}\pm} \approx \frac{2 l_D \Delta_L}{R_L^2} \frac{\beta_{\rm rec}}{\beta_{\rm wind}} \kappa_\pm \frac{c\Phi_{\rm mag}}{e},
\label{eq:capture}
\end{equation}
with $\kappa_\pm$ the multiplicity (multiplier of a fiducial Goldreich-Julian outflow rate $c\Phi_{\rm mag}/e$) that gives the number of {\it pairs} in the total outflow, $\beta_{\rm rec}$ the reconnection speed in units of $c$ and $\beta_{\rm wind}$ the polar wind outflow velocity (set equal to unity).  $\Delta_L$ is the thickness of the current channel at the light cylinder, assumed equal to the half height of the diffusion region, and $l_D$ is the length of the diffusion region. Pressure in the diffusion region expels the captured pairs from the diffusion region, outwards along the current sheet in the wind and inwards along the auroral current channels.  Expression (\ref{eq:capture}) assumes the plasma flux in the wind has a gradient across ${\boldsymbol B}$, as is likely since the accelerating electric field in the polar cap that leads to the pair creation that feeds the wind is small near the cap edge. The pressure supported electric field provides the accelerator which sorts the particles in the diffusion region into a precipitating beam (electrons in the geometry shown in Figure \ref{fig:current_system}) and an oppositely charged beam traveling outwards in the wind current sheet - the charge signs of the beams are as required by the global electrodynamics. The channel thickness is almost certainly comparable to the skin depth in the pairs. Taking the plasma gradient into account leads to the lower limit to the channel thickness
\begin{equation}
\Delta_{L,{\rm min}} = \frac{c}{\omega_{p\pm} [\delta = c/\omega_{p\pm}(\delta)]} \approx R_L \left(\frac{m_\pm c^2 \gamma_\pm}{2\kappa_\pm e \Phi_{\rm mag}} \right)^{1/3},
\label{eq:channel_width}
\end{equation}
where $\delta $ is the distance across $\mathbf{B}$ from the formal current sheet location, with a precipitating beam flux in the return current channel
\begin{equation}
F_\vee = \frac{l_D}{R_L} \kappa_\pm \frac{c\Phi_{\rm mag} /e}{2\pi R_L^2}\left(\frac{R_L}{r}\right)^3.
\label{eq:precip_flux}
\end{equation}
$\gamma_\pm$ is a measure of the flow momentum and of the comparable momentum dispersion of the polar plasma flow emerging from the inner magnetosphere, predicted by pair cascade models to be on the order of $10^2$. Numerically, (\ref{eq:channel_width}) yields a very small value, on the order of meters to hundreds of meters, the specific value depending on $\Phi_{\rm mag}$ and $\kappa_\pm$. The kinetics of relativistic reconnection being a largely untrodden subject, $l_D$ is more or less unknown - it could be as small as $\Delta_L$ itself (Petscheck style reconnection),  or as large as appears in the numerical dissipation  driven reconnection  observed in the force free  simulations, $l_D \sim 0.1 R_L$ (A. Spitkovsky, personal communication). In the terrestrial magnetosphere,  satellite observations suggest the diffusion region length is intermediate between the ion skin depth and the macroscopic scales.  For the discussion here (and in the more detailed report in preparation), treating $l_D$ as a parameter to be constrained by model comparisons to observations appears to be the wisest strategy. In analogy to observed non-relativistic reconnection, one expects $\beta_{\rm rec} \sim 0.1 v_{\rm Alfven}/c = 0.1 $, a value supported by the few PIC simulations of relativistic reconnection [{\it e.g.}, \cite{zenitani07}].

The precipitating particles form the ``hanging charge clouds'' invoked by \cite{gold69} to cause the electrostatic extraction of return current from the star's atmosphere, which happens if $n_\vee (R_*)  = F_\vee (R_*)/c \gg n_{GJ}(R_*)$, or , from (\ref{eq:precip_flux}), if 
\begin{equation}
\frac{l_D}{R_L} \frac{\beta_{\rm rec}}{\beta_{\rm wind}} \kappa_\pm \gg 1.
\label{eq:cloud}
\end{equation}
The condition in (\ref{eq:cloud}) is satisfied if $l_D$ is large compared to the skin depth, but still small compared to the numerical dissipation determined length observed in the force free simulations. It can be shown (\citealt{arons11}, and in preparation) that  the consequent minimum total potential drop in the twin beam channel is 
\begin{eqnarray}
\Delta \Phi_{\rm min} & \approx & -\frac{1}{8} \Phi_{\rm mag} \frac{R_*}{R_L} 
                       \left( \frac{m_\pm c^2 \gamma_\pm}{2e \Phi_{\rm mag} \kappa_\pm} 
                        \frac{\beta_{\rm wind}}{\beta_{\rm rec}} \right)^{1/3} 
                          \frac{l_D}{\Delta_L} \cos i 
                        \nonumber \\                       
                       & =& -\frac{1}{8} \Phi_{\rm mag} \frac{R_*}{R_L} 
                            \left(\frac{\beta_{\rm wind}}{\beta_{\rm rec}} \right)^{1/3}
			 \frac{l_D}{R_L}  \cos i .
\label{eq:channel_potential}
\end{eqnarray}
If $l_D/R_L$ really is as large as 0.1, rather than comparable to the minimum scale $\Delta_L$, the accelerating potential in the return current channel is more than large enough (around $10^{13}$ Volts, in the Crab and Vela pulsars) to drive curvature gamma ray emission from the beams, which limits the particle energies by radiation reaction. Such energies are high enough to lead to pair creation, which may limit the acceleration, although as is clear from expression (\ref{eq:genOhm}), parallel potential drops in the current carrying region can be sustained even if the plasma is dense.  

%If one models the gamma ray emission as being directly from the beams, one needs to take account of the electron beam going inwards, in acute geometry, where the upward beam is ions (generically, protons) - for obtuse geometry, the outbound beam is electrons, with an inbound positron beam.  Pair creation (through $\gamma - \gamma$ interactions with softer photons from the star), when it is important, makes the obtuse and acute geometries equivalent.

The high energy radiation might be synchrotron emission.  The counterstreaming beams are electromagnetically two stream and shear unstable.  Since $\omega_{p,beam}$ can be comparable to the relativistic cyclotron frequency in the outer magnetosphere, the growing waves can excite finite Larmor gyration of the particles in the current channel, thus producing incoherent emission through hard X-rays and gamma rays are possible. These X-rays are an alternative to soft photons from the star, as targets for $\gamma - \gamma$ pair production.  If the lower frequency waves can escape the plasma, they are a direct source of coherent emission, perhaps of interest to modeling giant radio pulses, which appear to come from the outer magnetosphere.

%\newpage
\vspace*{-2.8cm}

\subsection{Follow the Mass}

The Force-Free model says nothing about the nature of the plasma, that localizes $E_\parallel$ to a boundary layer in the magnetosphere and in the wind that emerges. Pulsar Wind Nebulae (PWNe) demonstrate that pulsars lose rest mass at a rate large compared to the fiducial electrodynamic particle loss rate $c\Phi_{\rm mag} /e$ [for a review of PWNe phenomenology, see \cite{gaensler06}]. The only known explanation is pair creation in the pulsars' magnetospheres. In many nebulae, the X-ray emitting particles rapidly lose energy to synchrotron radiation.  Then the nebulae are particle and energy calorimeters, allowing direct inference of the pair multiplicity in the wind, of TeV to PeV pairs.  The measured injection rates of X- and $\gamma$-ray emitting particles $\dot{N}_\pm$, up to $\sim 10^{38.5}$ pairs/s,  compare well to the predictions of existing pair creation models ({\it e.g.} \citealt{hibsch01}), yielding multiplicities $\kappa_\pm \equiv e\dot{N}_\pm /c \Phi_{mag}$ up to $\sim 10^4$. However, PWNe are also radio synchrotron emitters, radiation that samples much lower energy populations (100 MeV to 10 GeV), whose radiative efficiency is much less than their X-ray emitting cousins.  The result is a much larger population of pairs, whose radiative lifetime exceeds that of the nebulae.  The most efficient hypothesis is that these particles come from the embedded pulsars also, an idea supported by spectral continuity and by exotica, such as the observation of radio ``wisps'' near the Crab pulsars \citep{biet01}. Applying simple evolutionary models allows one to infer time averaged injection rates.  Early estimates \citet{shklovsky68} and more recent evaluations by {\it e.g.} \citet{atoyan96, slane10, bucc11} yield {\it lower limits} for multiplicities $\kappa_\pm$ all in excess of $10^5$  and {\it upper limits} for wind 4-velocities $\Gamma_{\rm wind} = \dot{E}_R /\dot{M}c^2 = e \Phi_{\rm mag}/2 \kappa_\pm m_\pm c^2$ all less than $10^5$ in a number of nebulae.  The data are the best for the younger systems, although even for these, the lack of far infrared data inhibits the analysis.\\
\newline
\begin{center}
\begin{tabular}{ccccc}
PWN Name & $\Phi_{\rm mag}$ (PV)  & Age (yr)  & $ \bar{\kappa}_\pm$  &  $\Gamma_{\rm wind}$ \\
 \hline
Crab &    100  & 955  &  $10^6$ & $5 \times 10^4$ \\
3C58 &  15  &  2100  & $ 10^{4.7}$ & $3 \times 10^4$ \\
B1509 & 120  & 1570 & $ 10^{5.3}$ & $1 \times 10^4$ \\
Kes 75 & 22 & 650 & $10^5$ & $7 \times 10^4$ \\
W44 & 1 & $20.3 \times 10^3$ & $ 10^5$ & $10^4 $ \\
K2/3 Kookaburra  & 5.5  & $13 \times 10^3$ & $10^5$ & $ 10^4$ \\
HESS J1640-465  &  3.5  & $  10^4 $ & $  10^6$ & $  10^{3.6}$
\end{tabular}
\end{center}

The inferred multiplicity excesses are a puzzle for pair creation theory, which have been apparent since \citet{shklovsky68} inferred a total injection rate of $10^{41}$ electrons/s needed to supply the radio emission of the Crab Nebula. These may be resolvable by appealing to magnetic anomalies near the neutron stars' surfaces, the simplest being an offset of the dipole center from the stellar center, which strengthens the magnetic field at one pole (\citealt{arons98, harding11}). The increase this gives to the magnetic opacity can be greatly enhanced if the magnetic axis is also tipped with respect to the radial direction, since then gravitational bending of photon orbits with respect to the $B$ field direction much increases the magnetic opacity for pair creation. Such phenomenological modifications of the low altitude magnetic field must respect the observation that radio beaming morphology is consistent with the magnetic field being that of a star centered dipole quite close to the star \citep{rankin90, kramer98}. This problem warrants quantitative investigation.

The large inferred multiplicities imply the wind 4-velocities $\Gamma_{\rm wind}$ to be small compared to the much quoted value of $10^6$ inferred by \cite{kennel84} in their model of the Crab Nebula's optical and harder emission. The large mass loading and the inferred low wind four velocity has a large impact on the much storied ``$\sigma$ problem'' of pulsar winds.  In ideal MHD, the ratio $\sigma \equiv B^2/8\pi m_\pm n_\pm \Gamma_{wind}c^2$ of magnetic energy to kinetic energy in the wind is conserved outside the fast magnetosonic radius (since for a cold flow the wind does not substantially accelerate outside this surface) and is large - even with the increased mass loading found from recent nebular studies, $\sigma$ is always well in excess of several hundred. Nevertheless the wind behaves at its termination shock as if $\sigma$  at that distance is small - MHD models of the nebulae suggest $\sigma$ at the termination shock is on the order of 0.02 in the Crab Nebula ({\it e.g} \citealt{delzanna04}) and similar values are plausible in other systems. 

\begin{figure}[H]
\begin{center}
%\unitlength = 0.0011\textwidth
\hspace{10\unitlength}
\begin{picture}(300,-110)(0,15)
\put(-20,-230){\makebox(300,200)[tl]{\includegraphics[width=5in]{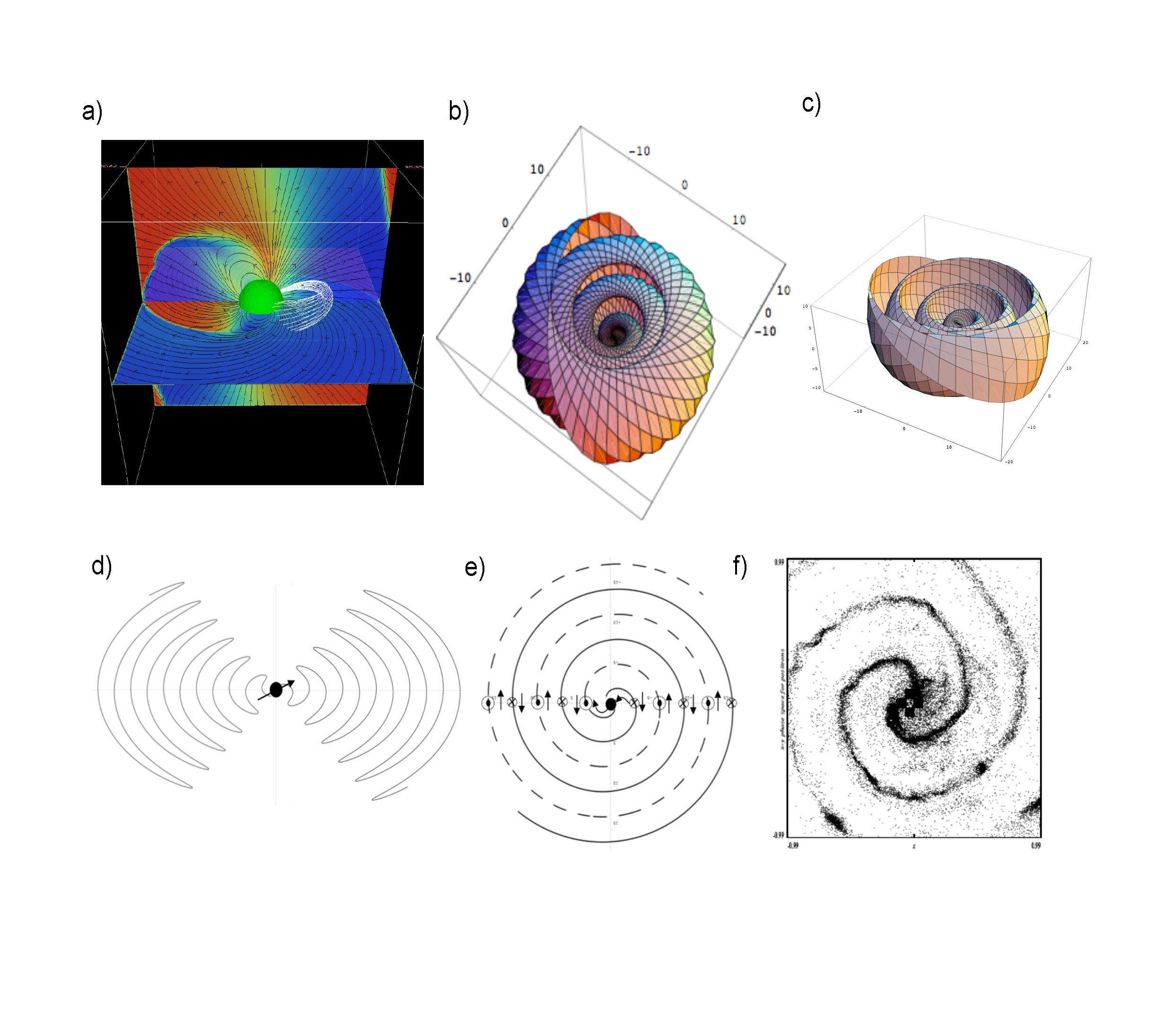}}}
\end{picture}
\end{center}
\vspace{10cm}
\caption{a) Magnetic Geometry of a Force-Free Rotator for $r<2R_L$, for $i = 60^\circ$,
from Spitkovsky (2006). The rapid transition to inclined split monopole field geometry for
$r > R_L$ is apparent. b) Geometry of the current sheet from the split monopole model for $i = 60^\circ$, $r > R_L$.  For clarity, only one of the two spirally wound current sheets is shown. As $i \rightarrow 90^\circ$, the sheets 
 almost completely enclose the star; for $r \gg R_L$, the spirals are tightly 
 wrapped ($B_r \ll B_\phi$) and the current sheet surfaces
 closely approximate nested spheres. 
  c) One sheet for $i=30^\circ$, shown for clarity. d) Meridional cross section
of the current sheet for $i=60^\circ$. e) Equatorial cross section snapshot of the current sheet, showing the two arm spiral form.  The arrows show the local directions of the 
magnetic field; the dots and crosses show the direction of the current flow. Panels b)-e) were constructed using Bogovalov's \citep{bogo99} analytic model of the asymptotic wind. f) Current sheet from a 2D PIC simulation of the inner wind, from unpublished work by Spitkovsky (used by permission),}
 \label{fig:stripes}
\end{figure}

\cite{coroniti90} suggested that because the wind of an oblique rotator has the magnetically striped structure shown in Figure \ref{fig:stripes}, magnetic dissipation of the corrugated B field, generically of a resistive nature propelled by instabilities of the current flow in the current sheet separating the stripes\footnote{This wrinkled current sheet, frozen into the wind,  is the continuation of the sheet separating the closed and open zones interior to the light cylinder, as is apparent in Figure \ref{fig:current_system}.} might destroy the magnetic field of the wind interior to the termination shock, thus converting a high $\sigma$ flow into a weakly unmagnetized plasma ($\sigma \ll 1$).  If the current sheets separating the magnetic stripes are to merge with a speed $v_s < c$ as measured in the proper frame of the flow, before they reach the termination shock located at distance $R_{TS}$ from the neutron star, the merger time in the PWN frame $T_{\rm merge} = \pi \Gamma_{\rm wind}^2 (R_L/v_s)$ must be less than the flow time to the termination shock $R_{TS}/c$, therefore $\Gamma_{\rm wind} < \sqrt{(R_{TS}/\pi R_L) (v_s/c)} = 5 \times 10^4 \sqrt{v_s/c} \; ({\rm Crab})$ must be satisfied if Coroniti's model is to be viable
-- see \citet{lyubarsky01}, \citet{kirk03} and \citet{arons08} for models with greater or lesser detail that address this issue.  This inequality is satisfied for multiplicities above $10^5$, which does appear to be the case for the young PWNe recently analyzed \citet{bucc11}. Possible mechanisms that can lead to the necessary dissipation are collisionless tearing and drift-kink instabilities of the current sheet, considered as if it were a flat sheet \citep{coroniti90, zenitani07} and an interesting Weibel-like instability due to interaction between the sheets \citep{arons08}, an effect strongest in the equatorial sector where the folded sheet appears locally as neighboring flat sheets with antiparallel current flow in the latitude direction.  

Such effects can be enhanced if the acceleration in the wind zone $R_L  < r < R_{TWS}$, driven by the pressure gradient in the wind, itself created by the dissipative heating of the wind \citep{lyubarsky01}, drives Rayleigh-Taylor instabilities of the current sheets \citep{lyub10}. If this occurs, the small scale disruptions of the current sheets can enhance the rate of magnetic decay, by shrinking the spatial scales over which the dissipation has to operate. A probable limitation of this idea is that it is formulated in such a way as to neglect the magnetic pressure gradient  between the center of each current sheet and the strongly magnetized stripes separating the sheets - each sheet is modeled as a uniform unmagnetized slab lying on top of the confining uniform magnetic field in the effective gravity felt in the frame co-moving with the accelerating wind.  Lyubarsky's chosen configuration is always unstable to incompressible ``fluting'' disturbances (in the sense of so-called flutes on a column), for sufficiently short wavelengths across $B$ and long wavelength parallel to $B$. However, as is well known \citep[and references therein]{arons76}, the real inhomogenous system, with magnetic pressure increasing from inside the plasma layers to the more rarefied and colder regions outside, is readily stabilized by the magnetic pressure gradient, against which potentially unstable ``heavy'' plasma mass elements have to do work as they try to fall into the low density magnetized region. Properly analyzing this instability requires a more sophisticated equilibrium model of the accelerating stripes and current sheets than that used in \citet{lyub10}.

Upon injection into the PWN, the flow decelerates at and beyond the wind "termination working surface" (TWS) that occurs where the the dynamic pressure of the wind $\dot{E}_R/4\pi R_{TWS}^2 c$ balances the pressure of the previously injected plasma and magnetic fields captured in the nebular bubble\footnote{I discuss here only the young high voltage nebulae, where pressure due to motion through the interstellar medium and to reverse shocks from the surrounding supernova remnant do not have major effects on the structure - these young systems are the most useful for probing the particle acceleration physics and the plasma properties.}. In the Crab Nebula, the TWS is usually identified with the prominent ring of emission seen in the Chandra image of the inner nebula modeled as a MHD shock wave, located at about the same location as the other signs of activity seen as indicating the termination of the free wind from the pulsar (``wisps'', etc., reviewed by \citealt{hester08}.)  Because the lack of identifiable emission from the region interior to the TWS suggests the upstream flow is cold, and (according to MHD flow models for the relativistic wind) is super-magnetosonic, the TWS is thought to be the simplest flow termination ``catastrophe'',  a relativistic magnetosonic shock wave ({\it e.g.} \citealt{kennel84}). Analogy to non-relativistic shocks in the supernova remnants, which clearly are efficient converters of flow energy to non-thermal, power-law-like spectra of accelerated particles ({\it e.g.} \citealt{reynolds11}), suggested to many that the non--thermal particle spectra inferred in the Crab and other PWNe arise from some form of shock acceleration. %The idea that the diffusive shock acceleration mechanism underlying the acceleration in shell SNR shocks is also behind the acceleration observed in the Crab's TWS figured prominently in the adoption of the shock model in the MHD models advanced with increasing sophistication for the structure of the inner Crab Nebula, believed to apply to other PWNe .

However, despite the sophistication achieved in MHD modeling of the inner Crab, especially of its variable wisp structures \citep{camus09}, no clear identification of the shock itself has appeared -- in particular, the Chandra ring has eluded the MHD modelers.  The applicability to other PWNe has also been unclear.  At a basic physical level, it is unlikely that diffusive shock acceleration (DSA), assumed 
\newpage 
\vspace*{-2.8cm}
\noindent by most to be the microphysics in the shock acceleration, actually is at work at the TWS in the Crab and other PWNe. The magnetic field in the wind is transverse to the flow - in the Crab, the field  winds up with $\sim 10^9$ turns between the light cylinder and the TWS.  It has long been known that DSA has difficulties in transverse shock geometry. In relativistic shocks, the process stalls because particles follow field lines, and as soon as the angle between the upstream field and the shock normal exceeds $1/\Gamma_1$, with $c\beta_1 \Gamma_1 \gg c$ the upstream 4-velocity, the velocity normal to the shock of a particle attempting to return to the upstream medium becomes less than the upstream fluid speed, so that bouncing between upstream and downstream becomes impossible \citep{begelman90} - the shocks are superluminal.  That difficulty can be overcome, if large amplitude, short wavelength  magnetic turbulence exists with amplitude sufficient to cause fast scattering across the {\it average} magnetic field \citep{niemic06, lemoine06, pelletier09}.  However, extensive PIC simulations of superluminal magnetosonic shocks in otherwise unstructured upstream media (no macroscopic upstream inhomogeneities, such as clumps in density, no current sheets, no particle component in the upstream flow with large Larmor radii at encounter with the shock,...) has shown that the shocks themselves generate no such turbulence  (\citealt{sironi09} and references therein, Spitkovsky \& Arons, in prepartstion), when the upstream magnetization $\sigma_1 \equiv (B^2/4\pi m n c^2 \Gamma)_{\rm upstream} $ is as large as is inferred in macroscopic MHD models of nebular surface brightness.

\subsection{Beyond MHD}

These and other problems have directed attention to extending the study of the TWS to conceptual realms larger than an MHD shock in an unstructured medium.  The essential consideration, driven by the angular anisotropy manifest in the nebular emission around many pulsars, is that there is something special about the angular sector of the nebula around the plane of the rotational equator of the underlying star - the X-ray torus and the Chandra ring in the Crab and the partial X-ray tori in other PWNe \citep{ng04, kargeltsev08} indicate the significance of this region.  The simplest special aspect follows from the angular dependence of the Poynting flux of a $\sigma \gg 1$ wind: $c ({\boldsymbol E} \times {\boldsymbol B}/4\pi ) \cdot {\boldsymbol e}_r =( \dot{E}_R /4\pi r^2 c) \cos^2 \lambda, \; \lambda =$ rotational latitude \citep{michel73}, a description that certainly covers the inner wind $r \ll R_{TWS}$.  MHD models of the Crab Nebula's post TWS flow that use this energy flux as input yield good models of the X-ray surface brightness, if two not unreasonable additional assumptions are made:  a) a particle flux uniform in latitude and b) an extra suppression of the energy flux toward the equatorial plane designed to represent hypothesized decay of the magnetic field upstream or at the TWS in the split monopole geometry of the outflow. For example, \citep{delzanna06} considered a total energy flux $\propto \cos^2 \lambda \tanh^2(\lambda/\lambda_0)$, with $\lambda_0$ the angular half-thickness of the putative current layer across which the wind's toroidal field reverses. They found  $\lambda_0 \sim 6^\circ$ and $\langle \sigma_1 \rangle \approx 0.02$ (the average is over $4\pi$ steradians) gives the best comparison of the synthetic surface brightness map to the Chandra X-ray images. Whether field decay occurs well upstream, as Coroniti envisioned, or occurs within the TWS itself \citep{lyubarsky03}, remains an unresolved issue. Field decay within the TWS is of possible importance if the upstream magnetic stripes do survive all the way to the TWS, for then reconnection, driven by the shock induced increase in the pressure, can dissipate the oppositely directed stripes of magnetic field \citep{lyubarsky03, sironi11}

\subsubsection{Current Starvation and Regeneration of Strong Vacuum-like Waves: Does it Happen?}

From a larger point of view, the non-trivial difficulties with the simplest variants of the standard model have, over the years, led various authors to suggest giving up MHD (even with dissipation) entirely.  A hoary motivation to do so, going back to \citet{usov75} and \citep{michel94}, is the claim that as the wind density drops $\propto r^{-2}$, current in the stripes becomes starved, since $J \propto r^{-1}$ so that at large enough radius, typically estimated to be $r \gtrsim 100\; {\rm AU} \approx 10^{15} \; {\rm cm} \ll R_{TWS}$, $J/2en_\pm \propto r$ exceeds $c$, an impossibility for conduction currents. This idea comes from simple use of Ampere's law in the proper frame of the flow, with ${\boldsymbol J}^\prime \propto {\boldsymbol \nabla}^\prime \times {\boldsymbol B}^\prime \propto 1/ R_L r \Gamma_{wind}^2$,  from ${\boldsymbol \nabla}^\prime = {\boldsymbol \nabla}/\Gamma_{wind}, \; {\boldsymbol \nabla} \sim 1/$stripe wavelength $=1/R_L $ and ${\boldsymbol B}^\prime = {\boldsymbol B}/\Gamma_{wind}$. The (usually unstated) assumption is that the relevant magnetic gradient scale is the stripe wavelength (\emph{e.g.}, \citealt{kundt80, melatos96}), with current starvation setting in for $R_{starve} < r \ll R_{TWS}$. The supposition then has been advanced that MHD must fail,  the frozen-in wave structure becomes a travelling wave ven in the proper frame of the plasma flow, and displacement current takes over from conduction current, thus resurrecting the strong, vacuum-like waves that were the foundation of pulsar rotational energy loss models in the early days of pulsar theory. Hypothesizing that this transition occurs rather abruptly, \citep{melatos98} derived jump conditions describing the transition from the frozen-in, comoving standing wave (in the wind proper frame) described by the striped wind model (a nonlinear entropy wave, as perceived from the point of view of MHD wave theory) to the superluminally propagating, strong EM waves conjectured in this idea; following the same idea, \citet{akra12} rediscovered and extended these jump conditions, as well as substantially extending the equilibrium theory of strong waves in a cold plasma.  

Such models have not shown that the hypothesized mode conversion actually can be realized - that would require an evolutionary or instability calculation, showing that the nonlinear entropy mode actually can couple to the superluminal strong electromagnetic waves; the existing models assume that coupling occurs and use jump conditions to characterize the supposed transition. Viewed from the perspective of small amplitude wave theory, that is a bit unlikely - the dispersion relation for the MHD entropy mode, $\omega = 0$ (the $\prime$ indicating plasma rest frame quantities is omitted for notational simplicity), never crosses that of the superluminal modes, except exactly at propagation cutoff, a non-propagating version of the wave where existing theory and simulation ({\it e.g.}, \citealt{leboeuf82, skaeraasen05} and references therein) suggests the waves are unstable and subject to strong damping. Nevertheless, the existence of even that limited resonance suggests that there might be merit in an evolutionary investigation, to see if the hypothesized transition might actually be realized -- then a more properly posed stability investigation would yield more useful information on the plasma heating and radiation characteristics that could be useful in employing this kind of model for studying relativistic striped winds in the real world.  A well chosen simulation is probably the best tool to study the question.

That idea that current starvation might occur and lead to a transition from MHD wind to some kind of superluminal electromagnetic wave was first made explicit by \citet{usov75}. His oft repeated  argument ({\it e.g.} \citealt{michel94}) may be ill founded, based as it is on the use of the stripe wavelength $R_L$ as the scale for the current flow supporting the stripes. In fact, the stripes are supported by the current in the current sheet, whose thickness $\Delta$ \emph{probably} is on the order of the formal Larmor radius $r_L = m_\pm c^2 \langle \gamma_\pm \rangle / eB$, where $\langle \gamma_\pm \rangle$ is the average Lorentz factor of the particles \emph{inside} a sheet and $B$ is the magnetic field just \emph{outside} the sheet \citep{michel94}.  Pressure equilibrium between the plasma in the current layer and the magnetic field outside yields $r_L = c/\sqrt{2} \omega_{p\pm}$ - the current thickness is on the order of the skin depth, indicating the relation of this structure to an electromagnetic pulse penetrating an overdense plasma. The magnetic field jumps across the current sheet from $B_\phi (= \Phi_{mag} /r)$ to $-B_\phi$. Integrating Ampere's law across the layer yields the surface current carried by the sheet $j = cB_\phi/2\pi$.  Since $j = J\Delta$, the volume current supporting the stripes is $J = cB_\phi / 2\pi \Delta$ - the current sheet thickness $\Delta$ replaces the wavelength $R_L$ that appeared in the conventional argument put forward by Usov and his successors. Assuming, as is plausible, that $\Delta$ is a factor $k$ times $r_L$, $ k \sim$ a few, I find $J \approx eB_\phi^2 2\pi k m_\pm c \langle \gamma_\pm \rangle \propto r^{-2}$ - $J/enc$ does \emph{not} decline as the wind expands, except to the degree that plasma heating in the current sheet causes $\langle \gamma_\pm \rangle$ to increase with $r$. According to the this estimate,  $J/enc = B_\phi^2 / 2\pi k m_\pm c^2 \langle \gamma_\pm \rangle n = B_\phi^2 /2\pi k P_\pm = 4/k < 1 $ if the current sheet is at least 4 formal Larmor radii thick. If the MHD model satisfies this constraint, the stripes' fate is determined by magnetic dissipation, as Coroniti hypothesized.  Whether that increase occurs requires properly accounting for the density in the current layer, which consumes the plasma in the stripes, and accounting for radiative cooling as well as the finding a proper model of the dissipative heating. Most likely,  \emph{if the wind was not current starved as it was launched, it never becomes so.}  But a proper determination awaits much more refined current layer models than have appeared so far. Since radiation losses offer the prospect of actually observing something, such modeling should be of more than idle theoretical interest.

Whether the dissipation does work fast enough to destroy the stripes interior to the TWS depends on physics not fully characterized so far. All that is certain is that if the mass flux is as small as that described by \citet{kennel84}, there isn't enough time to annihilate the stripes (\citealt{lyubarsky01}) no matter what the detailed microphysics may be, but as discussed above, incorporation of the total mass loading, including that required to feed the radio nebulae (explicitly ignored by Kennel \& Coroniti), does allow for reconnection velocities to be less than $c$, thus reopening the possible relevance of upstream dissipation.

\subsubsection{Observing the Wind}

This subject has an overly theological flavor, because of the lack of direct observations - the theoretical physics is interesting, but as with all non-linear plasma problems, making serious progress and filtering out the extraneous from the relevant ideas needs experiments - numerical, laboratory and telescopic observations. The upstream wind and its putative short-wavelength structure cries out for simulation and, if possible, experiment, perhaps possible in laser generated plasmas.  Even more important, detection and characterization of photons (or their lack, through absorption of radiation coming from the pulsar) that can be \emph{unambiguously} attributed to the wind would greatly advance the subject.  Inverse Compton scattering due to the wind's supposedly relativistic bulk flow may have sufficient luminosity to allow detection of inverse Compton \emph{emission} from the wind \citet{bogo00} at $r < 100 R_L$, perhaps distinguishable from brighter unpulsed nebular emission by its different spectral characteristics. 

The dissipation of the current sheets in the {\it inner} wind ($r \ll R_{TS}$, while responsible for only a small fraction of the magnetic destruction, might appear as \emph{pulsed} emission to the observer at the pulsar's rotation period.   The point is, if emission is confined to a sequence of thin layers separated radially by spacing $R_{sep}$ (= the light cylinder distance $R_L$ in the nominal striped wind model), moving out with speed $c\beta_{wind} = c - c/2\Gamma_{wind}^{2}, \; \Gamma_{wind} >> 1$, then so long as the emission at a given frequency has duration (in the pulsar frame) small compared to $(R_L/2 c) \Gamma_{wind}^2 = (P/4\pi) \Gamma_{wind}^2 $, the radiation from the sheets will not overlap and will arrive at the observer as a series of pulses occurring at the underlying pulsar's rotation frequency $P$.  This alternate to the usual lighthouse beaming model for pulsed emission, first alluded to by \citet{michel71} and more explicitly by \citet{arons79}, hasn't been explored much.  The lighthouse model for pulse formation, was adopted early on because of its simple explanation of the variation of pulsars' \emph{radio} polarization with rotation phase, an explanation that has survived all challenges, then later adapted to gamma ray emission.  Gamma rays are now clearly understood to arise at radii much larger than those for the radio emission, which comes from deep within the stars' magnetospheres.  The wind emission idea offers the promise of probing the otherwise unseen wind upstream of the TWS, so if such emission could be {\it unambiguously} identified, a lot could be be learned about the (experimentally) mysterious wind.  The upsurge of interest in current sheets in the wind that followed Coroniti's model for the wind's $B$ field dissipation has motivated some recent development of the idea of pulsed emission from the sheets, by Kirk, Petri and collaborators \citep{kirk02, petri05, petri09, petri11} as an explanation for {\it pulsed} optical, X-ray and gamma ray emission from the Crab and other pulsars, while \citet{aharonian12} have offered a variantt of the \citet{kirk02} model with inverse Compton radiation emitted at $r \sim 50 R_L$ as a model for 100 GeV pulsed emission recently discovered from the Crab pulsar \citep{aliu11, aleksic11}. That model should be compared to the inverse Compton scattering model with emission originating from near but within the light cylinder offered by \citet{lyutikov12}. 

So far all these interesting ideas have not found unique and incontrovertible use in modeling pulsar observations\footnote{A criticism that also can be leveled at the lighthouse model for high energy pulsed emission - while more popular, it's not clear that it is is unique.} - there is huge theoretical freedom in this kind of phenomenological modeling, and so far there have not been any observations which uniquely require wind emission as the model (although the $\varepsilon > 100$ GeV pulses from the Crab come close to demanding consideration of the wind, as do the \citet{bai10} lighthouse beaming models for the lower energy gamma ray pulses.) Parameterized model results of pulsed emission from the wind can be fit to observations of the light curves; however, so far they have had even less predictive predictive power than lighthouse  models, which rely on relativistic flow along $B$ rather than across $B$.   Lighthouse models for outer magnetosphere emission do a little better for the pulsed optical emission, in the few cases when it has been detected, it that the magnetic geometry employed doesn't require \emph{ad hoc} additions to the model, something that is required in the striped wind emission model -  getting the sweep of optical polarization in the Crab required adding an ad hoc meridional component of $B$ that rotates through the current sheet, a component not apparent in the force-free simulations of the inner wind.  But current sheets could, in principle, have such embedded rotating vectors. Not having been designed to study current sheet structure,  the existing numerical models may or may not offer a serious objection to this interesting phenomenological model.  More focused simulations would be useful. And in general, developing more self-consistent versions of these many interesting ideas would make comparisons to observations less subject to theorists' wiggling out of discrepancies by freely adding \emph{ad hoc} components to the models.

Detection of absorption and other radiation transfer phenomena in the light curves and spectra of the underlying pulsar, which backlights the wind, might offer a probe with less modeling freedom. There is a large literature exploring the influence of propagation effects, including absorption, on the formation of the complex light curves of radio pulsars, both as alternates to, and resolutions of, difficulties and issues in modeling the frequency dependent light curves solely in terms of emission and geometric beaming characteristics ({\it e.g.} \citealt{wang10} and references therein).  There has been little such effort in using the properties of photon transfer through the relativistic wind in the interpretation of pulsars' appearance, at any photon energy, although a few instances of such efforts can be found. For example, long ago, \citet{wilson78} used the transparency of the wind to induced Compton scattering of the pulsed radio emission as a means of setting a \emph{lower} limit on the wind's flow Lorentz factor in the inner wind they found $\Gamma_{wind} > 10^4$.  \citet{barnard86} showed that the polarization limiting radius for radio waves of short period pulsars might fall in the inner wind, where the $B$  field is toroidal and does not change its direction with pulsar rotation, thus plausibly explaining the lack of sweep of the Crab pulsar's radio polarization with rotation phase,  even though the optical pulse does show rotational sweep of the polarization - the temporal coincidence of the pulses suggests they arise at the same place, where the $B$ vector does rotate. These examples show that transfer of radio waves through the wind might create imprints on the observed emission which could be useful in diagnosing the wind - radio waves interact much more strongly with the wind's plasma than is the case for the higher frequency optical, X- and $\gamma$ rays.

\section{Flares: Analysis of Consequences \label{sec:flares-analysis}}

The recent discovery of gamma ray ``flares'' in the Crab Nebula's emission (the data are discussed in \S \ref{sec:flare_obs}) may have made the problems with the standard model more urgent.  Not only is it hard to see how DSA can succeed in a shocked flow with magnetic field transverse to the flow - it is hard to see how any accelerator with electric field less than the magnetic field, as is true of all models based on scattering of particles in magnetic turbulence, can account for the large amplitudes, short time scales and high photon energies of the flaring events, either with or without an embedded shock.   At photon energies above 100 MeV, the $0.1- 1$ PeV particles have very rapid radiative losses. The synchrotron cooling time for particles of energy $E = \gamma m_\pm c^2$ is (assuming a uniform distribution of pitch angles) 
\begin{equation}
T_{synch} = \frac{6\pi m_\pm c^2}{c \sigma_{Thomson} B^2 \gamma} =
    \frac{9}{4\gamma \Omega_{c\pm}} \frac{c}{r_e \Omega_{c\pm}} = \frac{28.4 {\rm hours}}{B_{milliGauss}^{3/2} \varepsilon_{GeV}^{1/2}}.
\label{eq:synch_time}
 \end{equation}
 Here $\Omega_{c\pm} \equiv eB/m_\pm c, \; \sigma_{Thomson} = (8\pi /3) r_e^2, \; r_e = e^2/m_\pm c^2$, and the synchrotron radiation equivalence of the observed photon energy $\varepsilon$ to the characteristic energy $\varepsilon_c = (3/2)\hbar \Omega_{c\pm} \gamma^2$ has been used to obtain $\gamma = \sqrt{(2/3)(\varepsilon / \hbar \Omega_{c\pm}) } = 7.6 \times 10^9 \sqrt{\varepsilon_{GeV} / B_{milliGauss}}$.  Rapid scattering requires deflection of the particles in the magnetic field, which can't happen any faster than the relativistic cyclotron time 
\begin{equation}
 T_{cyc} = 2\pi \gamma /\Omega_{c\pm} = 754 (E_{PeV}/ B_{milliGauss}) \; {\rm hours}:
\label{eq:cyctime}
\end{equation}
\noindent  the GeV photon emitting particles lose their energy within a few percent of one cyclotron orbit\footnote{This estimate assumes the fluctuating magnetic field causing scattering and the overall magnetic field causing the accelerated, radiating motion (which includes the fluctuating field) have the same magnitude, as is assumed in models which assume diffusion at the Bohm rate. }. Thus
\begin{equation}
\frac{T_{cyc}}{T_{synch}} = \frac{8\pi}{9}  \frac{\Omega_{c\pm} r_e}{c} \gamma^2 = 
    \frac{8\pi}{9}  \frac{\Omega_{c\pm} r_e}{c} \frac{(2/3)\varepsilon}{ \hbar \Omega_{c\pm}} = 
    \frac{16\pi}{27} \alpha_F \frac{\varepsilon}{m_\pm c^2} = 26.6 \varepsilon_{GeV},
\label{eq:cyc2cool}
\end{equation}
independent of the magnetic field strength. $\alpha_F$ is the fine structure constant. Radiation losses can occur in time short
compared to the relativistic cyclotron time since emission of the photons in question occurs without the completion of a full  gyration around the particle's guiding center - only a fraction $\sim \gamma^{-1}$ of the full Larmor orbit need be executed for a photon's emission. The time to radiate a photon is only the nonrelativistic cyclotron time $2\pi /\Omega_{c\pm}$, not $T_{cyc}$. 

Of course, in most astrophysical circumstances the radiation time \emph{is} longer than the gyration time - usually, the particle has to emit lots of photons before its energy changes substantially. For the particular parameters of relevance to the Crab flares, expression (\ref{eq:cyc2cool}) shows that the cooling time is longer than the gyration period for particles emitting at energies less than 30 MeV ($E < 200$ TeV).

The fact that particles emitting photons with energy above 100 MeV cannot complete full Larmor orbits makes traditional DSA, and also second order Fermi acceleration, in the TS region unlikely as the basic accelerator at these highest energies - these schemes have particles bouncing between major deflections in magnetic turbulence, with the bouncing caused by their non radiative Larmor gyration in the  up and downstream magnetic fluctuations. The bouncing is slow compared to the particles' Larmor times. Therefore, this mechanism cannot bring the particles up to the energies required  for the observed $\gamma$-ray emission. Independent of the magnetic field strength, accelerating electrons and positrons up to PeV energies requires an acceleration rate larger than the Larmor gyration rate.  

An alternate view of this issue comes from considering the maximum energy a particle can achieve in any accelerator in the face of synchrotron losses. The accelerator has an electric field $\cal E$, which may be laminar or stochastic, thus particles gain energy at the rate
\begin{equation}
\dot{\gamma} =  \frac{qce}{m_\pm c^2} {\cal E} = \Omega_{c\pm} \frac{{\cal E}}{B}, 
\end{equation}
and
\begin{equation}
T_{accel}  =  \frac{\gamma}{\dot{\gamma}} = \frac{\gamma}{\Omega_{c\pm}} \frac{B}{\cal{E}} 
                 =  \frac{T_{cyc}}{2\pi} \frac{B}{\cal{E}} . 
\end{equation}
If radiation losses occur at the synchrotron rate\footnote{That rate assumes only that the radius of curvature of a particle's orbit be the relativistic Larmor radius, not that particles actually complete full cyclotron orbits.} with loss time as in (\ref{eq:synch_time}), then
\begin{equation}
\frac{T_{accel}}{T_{synch}} = \frac{T_{cyc}}{2\pi T_{synch}}\frac{B}{\cal{E}}
     = \frac{4}{9} \frac{r_e \Omega_{c\pm}}{c} \gamma^2 \frac{B}{\cal{E}} 
     = \frac{8}{27}\frac{\varepsilon}{m_\pm c^2} \alpha_F \frac{B}{\cal{E}}.
\label{eq:accel2synch}
\end{equation}
As particles accelerate, the radiation loss time decreases, until at high enough energy, the radiative losses balance the energy gains and no further acceleration is possible.  Thus the radiated photon spectrum should show exponential roll-off above the characteristic energy $\varepsilon_{max}$ obtained by setting the ratio in (\ref{eq:accel2synch}) equal to unity,
\begin{equation}
\varepsilon_{max} = \frac{27}{8} \frac{m_\pm c^2}{\alpha_F}\frac{B}{\cal{E}}  = 236 \frac{B}{\cal{E}} \; {\rm MeV},
\label{eq:Emax}
\end{equation}
an extension of the long familiar estimate \citep{guilbert83, dejager96} to explicitly include the dependence on the accelerating electric field. Figures \ref{fig:thirty-threeMonths} and \ref{fig:LightCurves2011} show that the flare durations exceed the synchrotron cooling time, therefore the accelerator is ``on'' during a flare - the radiation is not a consequence of an impulse of acceleration followed by cooling with ${\cal E}$ switched off during the emission - and that the particle energies have reached the radiation reaction limit. 
 
The spectra of the biggest flare to date have exponential cutoff energies somewhat too high to be interpretable as coming from an accelerator with ${\cal E} < B$ - the cutoff energy exceeds (\ref{eq:Emax}) by a noticeable amount, a fact which has led to the suggestion \citep{uzdensky11, cerutti11} that the electric field is associated with magnetic reconnection in a region with magnetic topology similar to a sheet pinch or a cylindrical pinch, where ${\cal E} > B$ in the middle of the pinch.  This kind of configuration is known to lead to flaring acceleration in magnetic confinement devices (during ``disruptive instability''), in the solar corona (solar flares), and in planetary magnetospheres (terrestrial magnetospheric substorms), where particles  in the weak $B$ region in the middle of the pinch can undergo runaway acceleration in the electric field rather than drift motions in the crossed $\cal{E}$ and $B$ fields. However, this hint from the high cutoff energy might be mitigated if the source region undergoes relativistic bulk motion toward us during flare times - a Doppler boost factor $\sim 2$ would allow the cutoff energy in the plasma rest frame to be consistent with ${\cal E} < B$ in that frame.  Doppler boosts also relax the constraints of rapid time variability on the size of the region in question.  The short time scales of the flares - both duration and even shorter internal variations, as seen in Figure \ref{fig:LightCurves2011} -  may well be attributable to disruptive instability (such as tearing) of a current layer

However, a Doppler boost of this magnitude is not apparent in the regions where high angular resolution observations at lower photon energies have revealed variability.  The ``wisps'', the ``knot'', the ``anvil'', the inner regions of the ``jet'' all show apparent motions with speeds less than $0.5c$.  \citet{komiss11} suggested that relativistic flow from the part of the curved TWS that is tangent to the line of line of sight, modeled as a relativistic MHD shock wave (a very nice explanation of the nebular ``knot'' seen next to the pulsar put forth by \citealt{komiss03}), could be a flare site with such bulk Doppler boosting properties.  However, as modeled within the confines of MHD, the flow velocities aren't high enough to give enough Doppler boost.  Also, to date, high resolution radio \citep{lobanov11}, optical and infrared observations (C. Max and R. Romani, personal communications) of the knot have shown no changes temporally coincident with the gamma ray flares.  That lack of coincident change has been used to exclude the pulsar itself as the source of the flares; the same reasoning suggests the knot also is not the prime suspect. 

\citet{bykov11} take the more conservative view that all we are seeing is intermittency in the magnetized turbulence required in the traditional DSA model.  The short time flaring time scales and the long time between major flares are attributed to assumed intermittency in the turbulence, with luminosity increasing as magnetic field increases and the hardening of the spectrum the result of the increased emissivity at $\varepsilon_{max}$ causing the exponentially declining spectrum at higher energies to be visible at higher $\varepsilon$ than in quiescence - since radiation reaction limits the particle energies, the characteristic energy of exponential rolloff is independent of the field strength, therefore the shape of SED of the emission does not change during increases and decreases of $B$.  Aside from the difficulty of making DSA function in the face of the strong radiation losses, the SEDs of the April 201l flare look too hard to be readily explained by this idea - the increasing flux at 100-300 MeV seen in panels 7 and 8 of Figure \ref{fig:spectra2011flare} are not reproduced by the ``toy'' model spectra shown by \citet{bykov11}. Nevertheless, variable $B$ in the flare site surely contributes to the variability of the emission, on top of the impulsive behavior of the accelerator.

My personal bet is that the flare site is the equatorial current sheet in the immediate upstream ($r \lesssim R_{TWS}$) and Chandra Ring region of the wind and TWS, where the outflow velocity is still relativistic, and the current sheet separating the northern and southern hemispheres of the outflow is a permanent feature of the structure\footnote{The picture I favor is close to Coroniti's model of dissipation in the wind \citet{coroniti90}. The striped  magnetic structure launched from the pulsar (see Figure (\ref{fig:stripes}) decays interior to the TWS, probably at $r \sim 0.1 R_{TWS}$ (Arons, in preparation),  which is possible in the highly mass loaded wind in the Crab and other young pulsars.}.  Tearing in the reasonably flat equatorial current layer destroys azimuthal symmetry, leaving an array of radial current beams and reconnected islands, themselves subject to kinking, causing Doppler beaming to have highly variable directions, a plausible source of the observed variability. The long build up to major tearing disruption, seen in the Earth's magnetotail ({\it e.g.} \citealt{ohtani92}), may be the origin of the long time between flares - that buildup time is limited by the transit time from pulsar to TWS, $\sim$ months; minor tearing events, occurring quasi-continuously, may be the origin of the restless behavior of the light curve shown in Figure \ref{fig:thirty-threeMonths}. Variable beaming also suggests we don't see all the flares that occur. The radial reconnection electric field has a large length over which it can do its work, allowing particles to reach energies up  to a large fraction of the total wind voltage $\Phi_{mag}$ - this contrasts with the more limited acceleration possible in the tearing of the closely spaced current layers in the equatorial flow region, when the stripes are assumed to survive all the to the TWS, as in the model studied by \citet{sironi11}. The configuration is that of a linear accelerator (as in the earlier work on a similar, AGN jet motivated model by \citealt{larrabee03} and references therein), therefore radiation reaction limitations are reduced, especially for particles which stay well focussed within the current layer \citep{uzdensky11, cerutti11} - the current layer is relatively thick, with characteristic half width set by the latitudinal toroidal field gradient set by the imbalance in the magnetic flux in the neighboring stripes at latitudes off the exact mid-plane \citep{coroniti90}, thus longer particle residence in the current layer and higher voltage drops are possible than is the case where the current sheet forms oppositely directed current layers situated only $R_L$ apart from each other, as in the striped geometry\footnote{The focussing of particles toward the current sheet's center is a rediscovery of the focussing principle long known to accelerator physicists ({\it e.g.}, \citealt{courant58}). That mechanism can be used as a means to confine particles over distances $\sim 10^7 R_L$ = sheet spacing, a length required if voltages of a PetaVolt or more are to be accessed by the linear accelerator, only if the fields are smooth and properly ``designed'' to a degree extraordinary for an astrophysical configuration, especially when subject to the uncontrolled macroscopic gradients introduced by the instabilities of the current sheet.  The Sironi \& Spitkovsky simulations of the striped configuration show that when the formal Larmor radius of the particles in the sheets becomes somewhat larger than the spacing, the acceleration saturates, because particles have finite Larmor radius drifts out of the sheet's core in the ``messy'' fields of the unstable layer.}.  A configuration without the short wavelength $= R_L$ structure, such as the equatorial current layer remaining after upstream decay of the stripes, has a huge advantage as a geometry for reconnection powered runaway acceleration in the current layer.  The accelerated particles form beam dumps at the ends of the current filaments, perhaps providing a model for the hot spots observed on the Chandra ring. Because the particles are runaways, the particles form a beam - in the limit of a strictly 1D beam model, the particle distribution is mono--energetic, but in general, the spectrum should be very hard, perhaps even inverted in energy space, thus explaining both the appearance of a new very hard component in the gamma ray spectrum during a flare, and the negligible emission at photon energies below 100 MeV). This scenario suggests the emission region to be the outer wind, at angular separation from the pulsar $\sim 0.2^{\prime \prime} - 5^{\prime \prime}$, but with faint emission at energies below the VHE band that twinkles, because of variable Doppler beaming, That time variability suggests the value of continuous high sensitivity and high angular resolution monitoring in radio, optical/infrared and X-ray bands - arc second angular resolution in the VHE band is out of the question, unfortunately. A quantification of these ideas will emerge soon.\\

\noindent {\bf Acknowledgements} 

\vspace*{0.2in}
The work described here has been supported by NSF grant AST-0507813, NASA grants NNG06GJI08G and NNX09AU05G and DOE grant DE-FC02-06ER41453. I have benefitted from discussions with E. Amato, N. Bucciantini, C. Max, R. Romani, J. Scargle, A. Spitkovsky, A.Timokhin and D. Uzdensky.


\begin{thebibliography}{99.}

\bibitem[Abdo {\it et al.}(2010)]{abdo10a}
Abdo, A.A., {\it et al.} 2010, ApJ Supp, 187, 460 (First LAT Pulsar Catalog)

\bibitem[Abdo {\it et al.}(2011)]{abdo11}
Abdo, A.A., {\it et al.} 2011, Science, 331, 739

\bibitem[Aharonian {\it et al.}(2012)]{aharonian12}
Aharonian, F.A., Bogovalov, S. V. \& Khangulyan, D. 2012, Nature, 482, 507

\bibitem[Akra \& Kirk(2012)]{akra12}
Akra, I., \& Kirk, J. 2012, ApJ, 745, 108

\bibitem[Aliu {\it et al.}(2011)]{aliu11}
Aliu, E. {\it et al.} 2011, Science, 334, 69

\bibitem[Aleksic {\it et al.}(2011)]{aleksic11} 
Aleksic, J., {\it et al.}  2011, ApJ, 742, 43 

\bibitem[Anile(1990)]{anile90}
Anile, A.M. 1990, `Relativistic Fluids and Magneto-fluids' (Cambridge: Cambridge U. Press)

\bibitem[Arons \& Lea(1976)]{arons76}
Arons, J., \& Lea, S.M. 1976, ApJ, 207, 914; 210, 792

\bibitem[Arons(1979)]{arons79}
Arons, J. 1979, Space Science Reviews, 24, 437

\bibitem[Arons(1981a)]{arons81a}
Arons, J. 1981a, in {\it Pulsars}, W. Sieber \& R. Wielebinski, eds. (Dordrecht: Reidel), 75

\bibitem[Arons(1981b)]{arons81b}
Arons, J. 1981b, Ap.J., 248, 1099

\bibitem[Arons(1983a)]{arons83a}
Arons, J. 1983a, ApJ, 266, 241

\bibitem[Arons(1983b)]{arons83b}
Arons, J. 1983b, in `Positron-electron Pairs in Astrophysics', R. Ramaty \& A.K. Harding, eds., AIPC vol. 101 (New York: American Institute of Physics), 163

\bibitem[Arons(1996)]{arons96}
Arons, J. 1996, A \& A Supp., 120, 49

\bibitem[Arons(1998)]{arons98}
Arons, J. 1998, in `Neutron Stars and Pulsars : Thirty Years after the Discovery', N. Shibazaki {\it et al.}, eds. (Tokyo, Japan : Universal Academy Press, Frontier Science Series  No. 24), 339

\bibitem[Arons(2008)]{arons08}
Arons, J. 2008, Int. J. Mod. Phys. D, 17, 1419

\bibitem[Arons(2011)]{arons11}
Arons, J. 2011, in `High-Energy Emission from Pulsars and their Systems', N. Rea and D. Torres, eds. Astrophys. \& Space
 Sci. Proceedings DOI: 10.1007/978-3-642-17251-9 (Berlin \& Heidelberg: Springer), 165
 
 \bibitem[Atoyan \& Aharonian(1996)]{atoyan96}
 Atoyan, A.M., \& Aharonian, F.A. 1996, MNRAS, 278, 525

\bibitem[Balbo {\it et al.}(2011)]{balbo11}
Balbo, M., Walter, R., Ferrigno, C., \& Bordas, P. 2011, A\&A, 527, L4

\bibitem[Bai \& Spitkovsky(2010)]{bai10}
Bai, X. \& Spitkovsky, A. 2010, ApJ,  715, 1282

\bibitem[Barnard(1986)]{barnard86}
Barnard, J.J. 1986, ApJ, 303, 280

\bibitem[Begelman {\it et al.}(1984)]{begelman84}
Begelman, M.C., Blandford, R.D., and Rees, M.J. 1984, Rev. Mod. Phys., 56, 255

\bibitem[Begelman \& Kirk(1990)]{begelman90}
Begelman, M.C., \& Kirk, J.G. 1990, Ap.J., 353, 66

\bibitem[Bietenholz {\it et al.}(2001)]{biet01}
Bietenholz, M.F., Frail, D.A., \& Hester, J.J. 2001, ApJ, 560, 254

\bibitem[Bietenholz {\it et al.}(2004)]{biet04}
Bietenholz, M.F., Hester, J.J., Frail, D.A., \&  Bartel, N. 2004, ApJ, 615, 794

\bibitem[Blandford \& Romani(1988)]{blandford88}
Blandford, R., \& Romani, R. 1988, MNRAS, 234, 57

\bibitem[Bogdanov {\it et al.}(2007)]{bogdanov07}
Bogdanov, S., Rybicki, G., and Grindlay, J. 2007, ApJ, 670, 668

\bibitem[Bogovalov(1999)]{bogo99}
Bogovalov, S.V. 1999, A \& A, 349, 1017

\bibitem[Bogovalov \& Aharonian(2000)]{bogo00}
Bogovalov, S.V., \& Aharonian, F.A. 2000, MNRAS, 313, 504

\bibitem[Bucciantini {\it et al.}(2006)]{bucc06}
Bucciantini, N., Thompson, T.A., Arons, J., Quataert, E., and del Zanna, L. 2006, MNRAS, 368, 1717 (arXiv astro-ph/0602475)

\bibitem[Bucciantini {\it et al.}(2011)]{bucc11}
Bucciantini, N., Arons, J., and Amato, E. 2011, MNRAS, 410, 381 (arXiv 1005.1831)

\bibitem[Buehler {\it et al.}(2012)]{buehler11}
Buehler, R., Scargle, J.D., Blandford, R., {\it et al.} 2012, ApJ, 749, 26 (arXiv 1112.1979)

\bibitem[Bykov {\it et al.}(2011)]{bykov11}
Bykov, A., Pavlov, G.G., Artemyev, A.V., \& Uvarov, Yu. A. 2011, arXiv 1112.3114 (MNRAS, in press)

\bibitem[Camus {\it et al.}(2009)]{camus09}
Camus, N., Komissarov, S., Bucciantini, N., \& Hughes, P. 2009, MNRAS, 400, 1241

\bibitem[Carlson, Langer \& Shaw(1994)]{langer94}
Carlson, J.M., Langer, J.S., \& Shaw, B.E. 1994, Rev. Mod. Phys., 66, 657-671

\bibitem[Cheng, Ho \& Ruderman(1986)]{ruderman86}
Cheng, K.S., Ho, C., \& Ruderman, M.A. 1986, ApJ, 300, 500

\bibitem[Cheng(1987)]{cheng87}
Cheng, K.S. 1987, Ap.J., 321, 799

\bibitem[Chung \& Melatos(2011)]{chung11}
Chung, C.T.Y., \& Melatos, A. 2011, MNRAS, 411, 2471

\bibitem[Cerutti {\it et al.}(2012)]{cerutti11}
Cerutti, B. and Uzdensky, D. A. and Begelman, M. C. 2012, ApJ, 746, 148 (arXiv 1110.0557)

\bibitem[Contopoulos {\it et al.}(1999)]{contop99}
Contopoulos, I., Kazanas, D., and Fendt, C. 1999, ApJ, 511, 351  

\bibitem[Contopoulos \& Spitkovsky(2006)]{contop06}
Contopulos, I., \& Spitkovsky, A. 2006, Ap.J., 643, 1139 

\bibitem[Coroniti(1990)]{coroniti90}
Coroniti, F.V. 1990, ApJ, 349, 538

\bibitem[Courant \& Snyder(1958)]{courant58}
Courant, E.D., \& Snyder, H.S. 1958, Ann. Phys., 3, 1

\bibitem[Del Zanna {\it et al.}(2006)]{delzanna06}
Del Zanna, L., Volpi, D., Amato, E., \& Bucciantini, N. 2006, A \& A, 453, 621

\bibitem[de Jager {\it et al.}(1996)]{dejager96}
de Jager, O. C., et al. 1996, ApJ, 457, 253

\bibitem[Deutsch(1955)]{deutsch55}
Deutsch, A. 1955, Ann. d'Ap., 18,1 

\bibitem[DelZanna {\it et al.}(2004)]{delzanna04}
DelZanna, L., Amato, E., and Bucciantini, N. 2004, A \& A, 421, 1063

\bibitem[Gaensler \& Slane(2006)]{gaensler06}
Gaensler, B., and Slane, P. 2006, Ann. Rev. Astron. Astrophys., 44, 17 

\bibitem[Goldreich \& Julian(1969)]{gold69}
Goldreich, P., and Julian, W.H. 1969, ApJ, 157, 869 

\bibitem[Gruzinov(2005)]{gruzinov05}
Gruzinov, A. 2005, Phys. Rev. Lett., 94, 021101

\bibitem[Gruzinov(2008)]{gruzinov08}
 Gruzinov, A. 2008, J. Cosmology \& Astroparticle Physics., 11, 2
 
\bibitem[Guilbert {it et al.}(1983)]{guilbert83}
Guilbert, P. W., Fabian, A. C., \& Rees, M. J. 1983, MNRAS, 205, 593
 
\bibitem[Harding(1981)]{harding81}
Harding, A.K. 1981, ApJ, 425, 267

\bibitem[Harding \& Muslimov(2002)]{harding02}
Harding, A.K., \& Muslimov, A.G. 2002, ApJ, 568, 862

\bibitem[Harding \& Muslimov(2011a\&b)]{harding11}
Harding, A.K., \& Muslimov, A.G. 2011a, ApJ Lett, 726, L10; 2011b, ApJ, 743, 181

\bibitem[Helfand {\it et al.}(1980)]{helfand80}
Helfand, D.J., Taylor, J.H., Backus, P.R., and Cordes, J.M. 1980, Ap.J., 237, 206

\bibitem[Hester(2008)]{hester08}
Hester, J.J. 2008, Annu. Rev. Astron. Astrophys., 46, 127Ð55

\bibitem[Hibschman \& Arons(2001)]{hibsch01}
Hibschman, J.A., and Arons, J. 2001, ApJ, 554, 624

\bibitem[Hoyle {\it et al.}(1964)]{hoyle64}
Hoyle, F., Narlikar, J.V., \& Wheeler, J.A. 1964, Nature, 203, 914

\bibitem[Jackson(1962)]{jackson62}
Jackson, J.D. 1962, `Classical Electrodynamics' (New York: Wiley), section 14.6

\bibitem[Kalapotharakos \& Contopoulos(2009)]{kala09}
Kalapotharakos, C., \& Contopoulos, I. 2009, A. \& A., 496, 495

\bibitem[Kargeltsev \& Pavlov(2008)]{kargeltsev08}
Kargeltsev, O., \& Pavlov, G.G. 2008, in `40 Years of Pulsars--Millisecond Pulsars, Magnetars and More', C.G. Bassa, Z. Wang, A. Cumming \& V. Kaspi, eds,
AIP Conference Proceedings No. 983, (New York: AIP), 171-185

\bibitem[Kaspi(2010)]{kaspi10}
Kaspi, V. 2010, PNAS, 107, 7147  (arXiv 1005.0876) 

\bibitem[Kennel \& Coroniti(1984)]{kennel84}
Kennel, C.F., \& Coroniti, F.V. 1984, ApJ, 283, 694 \& 710

\bibitem[Kirk {\it et al.}(2002)]{kirk02} 
Kirk, J.G., Skj{\ae}raasen, O.  \&  Gallant, Y.A. 2002, A\&A, 388, L32

\bibitem[Kirk \& Skj{\ae}raasen(2003)]{kirk03} 
Kirk, J.G. \& Skj{\ae}raasen, O. 2003, ApJ 591,366

\bibitem[Komissarov \& Lyubarsky(2003)]{komiss03}
Komissarov S. S., Lyubarsky Y. E., 2003, MNRAS, 344, L93

\bibitem[Komissarov(2006)]{komiss06}
Komissarov, S. 2006, MNRAS, 367, 19

\bibitem[Komissarov \& Lyutikov(2011)]{komiss11}
Komissarov, S. S. and Lyutikov, M. 2011, MNRAS, 414, 2011

\bibitem[Kotera \& Olinto(2010)]{kotera10}
Kotera, K., \& Olinto, A. 2010, Ann. Rev. Astron. Ap.,  49, 1

\bibitem[Kramer {\it et al.}(1998)]{kramer98}
Kramer, M., Xilouris, K.M., Lorimer, D.R., {\it et al.} 1998, ApJ, 501, 270

\bibitem[Kundt \& Krotscheck(1980)]{kundt80}
Kundt, W., \& Krotscheck, E. 1980, A\&A, 83, 1

\bibitem[Lampland(1921)]{lampland21}
Lampland, C.O. 1921, Publ. Ast. Soc. Pacific., 33, 79

\bibitem[Larrabee, Lovelace \& Romanova(2003)]{larrabee03}
Larrabee, D.A., Lovelace, R.V.E., \& Romanova, M.M. 2003, ApJ, 586, 72

\bibitem[Leboeuf {\it et al.}(1982)]{leboeuf82}
Leboeuf, J.N., Ashour-Abdalla, M., Tajima, T., {\it et al.} 1982, Phys. Rev. A, 25, 1023

\bibitem[Lemoine {\it et al.}(2006)]{lemoine06}
Lemoine, M., Pelletier, G. \& Revenu, B. 2006, Ap.J., 645, L49

\bibitem[Livingstone {\it et al.}(2007)]{livingstone07}
Livingstone, M., Kaspi, V., Gavrill, F., {\it et al.} 2007, Astrophys. Space Sci., 308, 317

\bibitem[Lobanov {\it et al.}(2011)]{lobanov11}
Lobanov, A.P., Horns, D., \& Muxlow, T.W.B. 2011, A \& A, 533, A10

\bibitem[Lyne {\it et al.}(2010)]{lyne10}
Lyne, A., Hobbs, G., Kramer, M., Stairs, I., \& Stappers, B. 2010, Science, 329, 408 

\bibitem[Lyubarsky \& Kirk(2001)]{lyubarsky01}
Lyubarsky Y. \& Kirk J.G. 2001, ApJ, 547, 437 

\bibitem[Lyubarsky(2003)]{lyubarsky03}
Lyubarsky, Y. 2003, MNRAS, 345, 153

\bibitem[Lyubarsky(2010)]{lyub10}
Lyubarsky, Y. 2010, ApJ letters, 725, L234

\bibitem[Lyutikov {\it et al.}(2012)]{lyutikov12}
Lyutikov,  M., Otte, N. \& McCann, A. 2012, ApJ, submitted (arXiv 1108.3824)

\bibitem[Medin \& Lai(2010)]{medin10}
Medin, Z., \& Lai, D. 2010, MNRAS, 406,1379 (arXiv 1001.2365)

\bibitem[Melatos \& Melrose(1996)]{melatos96}
Melatos, A., \& Melrose, D. 1996, MNRAS, 279, 1168

\bibitem[Melatos(1998)]{melatos98}
Melatos, A. 1998, Memorie della Societˆ Astronomia Italiana, 69, 1009

\bibitem[Michel(1969)]{michel69}
Michel, C.F. 1969, ApJ, 158, 727

\bibitem[Michel(1971)]{michel71}
Michel, F.C. 1971, Comments on Astrophysics and Space Physics, 3, 80

\bibitem[Michel(1973)]{michel73}
Michel, C.F. 1973, ApJ, 180, L133
 
\bibitem[Michel(1975)]{michel75}
Michel, C.F. 1975, ApJ, 197, 193 

\bibitem[Michel(1994)]{michel94}
Michel, F.C. 1994, Ap.J., 431, 397

\bibitem[Ng \& Romani(2004)]{ng04}
Ng, C.Y., \& Romani, R. 2004, ApJ, 601, 479

\bibitem[Niemic \& Ostrowski(2006)]{niemic06}
Niemic, J., and Ostrowski, M. 2006, Ap.J., 641, 982

\bibitem[Ohtani {\it et al.}(1992)]{ohtani92}
Ohtani, S., Takahashi, K., Zanetti, L., {\it et al.} 1992, JGR, 97, 19311

\bibitem[Ostriker \& Gunn(1969)]{ostriker69}
Ostriker, J., \& Gunn, J.\ 1969, ApJ, 157, 139 

\bibitem[Pacini(1967)]{pacini67}
Pacini, F. 1967, Nature, 434, 1107

\bibitem[Paschmann {\it et al.}(2002)]{pasch02}
Paschmann, G, Haaland, S., \& Treumann, R. 2002, `Auroral Plasma Physics' , Space Science Reviews,  103, 1-485

\bibitem[Pelletier {\it et al.}(2009)]{pelletier09}
Pelletier, G., Lemoine, M., \& Marcowith, A. 2009, MNRAS, 393, 587

\bibitem[Petri \& Kirk(2005)]{petri05}
Petri, J. \& Kirk, J.G. 2005, ApJ, 627, 37

\bibitem[Petri(2009)]{petri09}
Petri, J. 2009, A\&A, 503, 13

\bibitem[Petri(2011)]{petri11}
Petri, J. 2011, MNRAS, 412, 1870

\bibitem[Rankin(1990)]{rankin90}
Rankin, J. 1990, ApJ, 352, 247

\bibitem[Ray \& Saz Parkinson(2011)]{ray11}
Ray, P., \& Saz Parkinson, P.M. 2011, `High-Energy Emission from Pulsars and their Systems', N. Rea \& D.F. Torres, eds. (Heidelberg: Springer), 37

\bibitem[Rees \& Gunn(1974)]{rees74}
Rees, M.J., \& Gunn, J.E. 1974, MNRAS, 167, 1

\bibitem[Rees(1984)]{rees84}
Rees, M.J. 1984, Ann. Rev. Astron. Astrophys., 22, 471

\bibitem[Reynolds(2011)]{reynolds11}
Reynolds, S. 2011, Astrophys. \& Space Sci., 336, 257

\bibitem[Scargle(1969)]{scargle69}
Scargle, J.D. 1969, ApJ, 156, 401

\bibitem[Scott {\it et al.}(2003)]{scott03}
Scott, D.M., Finger, M.H., \& Wilson, C.A. 2003, MNRAS, 344, 412

\bibitem[Scharlemann {\it et. al.}(1978)]{saf78}
Scharlemann, E.T., Arons, J., and Fawley, W.M. 1978, ApJ, 222, 297

\bibitem[Shklovsky(1968)]{shklovsky68}
Shklovsky, I. 1968, `Supernovae' (New York: Wiley)

\bibitem[Slane {\it et al.}(2010)]{slane10}
Slane, P., Castro, D., Funk, S., {\it et al.} 2010, ApJ, 720, 266

\bibitem[Sironi \& Spitkovsky(2009)]{sironi09}
Sironi, L., \& Spitkovsky, A. 2009, Ap.J., 698, 1523

\bibitem[Sironi \& Spitkovsky(2011)]{sironi11}
Sironi, L., \& Spitkovsky, A. 2011, Ap.J., 726, 75

\bibitem[Sk{\ae}raasen {\it et al.}(2005)]{skaeraasen05}
Sk{\ae}raasen, O., Melatos, A., \& Spitkovsky, A. 2005, ApJ, 634, 542

\bibitem[Spitkovsky \& Arons(2004)]{spit04}
Spitkovsky, A., \& Arons, J. 2004, ApJ, 603, 669 

\bibitem[Spitkovsky(2006)]{spit06}
Spitkovsky, A. 2006, ApJ., 648, L51

\bibitem[Sturrock(1971)]{sturrock71}
Sturrock, P.A. 1971, ApJ, 164, 529 

\bibitem[Tavani {it et al.}(2011)]{tavani11}
Tavani, M., Bulgarelli, A., Vittorini, V., {\it et al.} 2011, Science, 331, 736

\bibitem[Timokhin(2006)]{timokhin06}
Timokhin, A. 2006, MNRAS, 368, 1055

\bibitem[Usov(1975)]{usov75}
Usov, V.V. 1975, Ap. \& Space Sci., 32, 375

\bibitem[Uzdensky \& Kulsrud(1997)]{uzdensky97}
Uzdensky, D.A., \& Kulsrud, R.M. 1997, Phys. Plasmas, 4, 3960

\bibitem[Uzdensky {\it et al.}(2011)]{uzdensky11}
Uzdensky, D. A., Cerutti, B. and Begelman, M. C. 2011, ApJ(Letters), 737, L40 (arXiv 1105.0942)

\bibitem[Wang, Lai \& Han(2010)]{wang10}
Wang, C., Lai, D. \& Han, J. 2010, MNRAS, 403, 569

\bibitem[Weisskopf {\it et al.}(2000)]{weisskopf00}
Weisskopf, M.C., Hester, J.J., Tennant, A.F., {\it et al.} 2000, ApJ, 536, L81

\bibitem[Wilson \& Rees(1978)]{wilson78}
Wilson, D.B., \& Rees, M.J. 1978, MNRAS, 185, 297

\bibitem[Zavlin \& Pavlov(1998)]{zavlin98}
Zavlin, V.E., \& Pavlov, G. 1998, A. \& A., 329, 583

\bibitem[Zenitani \& Hoshino(2007)]{zenitani07}
 Zenitani, S., \& Hoshino, M. 2007, ApJ, 670, 702

\end{thebibliography}
\end{document}